\documentstyle[12pt,epsf]{article}
\begin{document}
\hoffset=-1.2cm
\baselineskip=7.5mm
\hsize=16cm
\vsize=24cm
\begin{titlepage}
\begin{flushright}
DTP-95/06\\
February, 1995\\
\end{flushright}
{\centerline{\bf {ONE LOOP QED VERTEX IN ANY COVARIANT GAUGE:}}}
\vskip 2mm
{\centerline{\bf {ITS COMPLETE ANALYTIC FORM}}}
\vskip 2cm
\baselineskip=7mm
{\centerline{\bf{A. K{\i}z{\i}lers\"{u}{$^{1,2}$}, M. Reenders${^3}$ and
M.R. Pennington$^1$}}}
\vskip 5mm
{\centerline{{$^1$}Centre for Particle Theory,}}
{\centerline{University of Durham}}
{\centerline{Durham DH1 3LE, U.K.}}
\vskip 1cm
{\centerline{{$^2$}University of Istanbul,}}
{\centerline{Faculty of Science, Physics Department,}}
{\centerline{Beyaz{\i}t, Istanbul, Turkey}}
\vskip 1cm
{\centerline{{$^3$}Institute for Theoretical Physics,}}
{\centerline{Nijenborgh 4, NL-9747 AG}}
{\centerline{Groningen, The Netherlands}}
\vskip 2cm
{\centerline {ABSTRACT}}
{\noindent The one loop vertex in QED is calculated in arbitrary covariant
gauges as an analytic function of its momenta. The vertex is decomposed
into a longitudinal part, that is fully responsible for ensuring the Ward
and Ward-Takahashi identities are satisfied, and a transverse part. The
transverse part is decomposed into 8 independent components each being
separately free of kinematic
singularities in $\bf any$ covariant gauge in a basis that modifies that
proposed by Ball and Chiu. Analytic expressions for all 11
components of the ${O(\alpha)}$ vertex are given explicitly in terms of
 elementary functions and one Spence function. These results greatly
simplify in particular kinematic regimes.}
\end{titlepage}
\vfil\eject
\vskip 2cm
\section{Introduction}
\baselineskip=7mm
%
%
%
%
%
%
%
%
%
%
%
%
%
%
%
\indent
     This paper presents the calculation of the one loop vertex in QED in an
arbitrary covariant gauge. Why should we want to compute this? Needless to say,
it is the interactions that determine the structure and properties of
any theory. In QED the fermion-boson vertex is this basic interaction. Many, if
not most, physical phenomena are controlled by kinematic regimes in which the
interactions become strong. This determines, for instance, the spectrum of
hadrons and the
nature of confinement in QCD or the existence of ${e^+e^-}$ bound states
in the strong electromagnetic fields of heavy nuclei. Such phenomena can only
be studied by non-perturbative techniques using (as appropriate) the
Schwinger-Dyson or Bethe-Salpeter equations in the continuum or on the
lattice. In undertaking  studies of the non-perturbative nature of
gauge theories~\cite{Ref}, we immediately have to confront the issue of what is
the non-perturbative form of the fundamental fermion-boson interaction.
$Ans${\H{a}}$tze$ for this are needed to accomplish a truncation of the
heirarchy
of the field equations that are the Schwinger-Dyson equations. It is known
that the much used ${\it rainbow}$ approximation with its bare vertex,
${\gamma^{\mu}}$, while seductively simple, fails to respect the gauge
invariance and multiplicative renormalizability so crucial in determining
the structure of the theory  and the characteristics of observables. Thus
one must seek more sophisticated ans\"atze that do respect these key
properties.

       The only truncation of the complete set of Schwinger-Dyson
equations, that we know of, that maintains the gauge invariance and
multiplicative renormalizability of a gauge theory at every level of
approximation is perturbation theory. Physically meaningful solutions
of the Schwinger-Dyson equations must agree with perturbative results
in the weak coupling regime. Perturbation theory can thus serve as a guide
to allowed non-perturbative forms. To be concrete, we know that the complete
fermion propagator, ${S_{F}}$, of momentum ${p}$ involves two functions of
${p^2}$. This follows from the spin structure of the fermion propagator.
These two can be chosen to be ${F(p^2)}$, the wavefunction renormalization,
and ${M(p^2)}$, the mass function, so that
\begin{eqnarray}
{\it i}{\it S}_{F}(p)={\it i}\frac{F(p^2)}{\not\! p-M(p^2)}\qquad.
\end{eqnarray}
[This can be  (and is often) written in a variety of other ways, e.g.
${{\it S}_{F}(p)^{-1}=\alpha(p^2)\not\! p + \beta(p^2)}$, etc,
always involving two independent scalar functions.]
Since ${S_F(p)}$ is a gauge-variant quantity, these functions
${F(p^2)}$, ${M(p^2)}$ will in general depend on the gauge. They can be
calculated, in principle, at each order in perturbation theory.
At lowest order ${F(p^2)=1}$, ${M(p^2)=m}$, the bare mass. Now these
same functions must also occur in the fermion-boson vertex, since
the Ward-Takahashi identity relates the 3-point Green's function to the
fermion propagator in a well-known way. This is satisfied at every
order of perturbation theory. Indeed, such identities are true
non-perturbatively. Thanks to the work of Ball and Chiu~\cite{BC} we
know how to express the non-perturbative structure of the part of the vertex
(~a part conventionally called the  longitudinal component~) that
fulfills the Ward-Takahashi
identity in terms of the two non-perturbative
functions describing the fermion propagator. We have also learnt that
multiplicative renormalizability of the fermion propagator imposes
further constraints on the vertex but these have yet to be fully exploited,
though a start has been made~\cite{Gauget,CP,BP}. While the bare fermion-boson
vertex in a
minimal coupling gauge theory is simply ${\gamma^{\mu}}$, in general
the vertex involves twelve spin amplitudes that can be constructed from
${\gamma^{\mu}}$ and the two independent 4-momenta at the vertex as
elucidated by Bernstein~\cite{Currents}. This would suggest that the complete
fermion-boson vertex involved a large number of independent functions.
However, some of these at least must be related to the fermion
functions ${F(p^2)}$, ${M(p^2)}$, not to mention the analogous
boson renormalization function ${G(p^2)}$. It is to the nature of
these forms that perturbation theory can be a guide, but only if we calculate
in an arbitrary gauge. For instance, if we calculated the vertex in massless
QED merely in the Landau gauge  we would find  the ${\gamma^{\mu}}$ component
was like its bare form just ${\gamma^{\mu}}$. This would serve little as a
pointer to the form
${\frac{1}{2}\left[F^{-1}(k^2)+F^{-1}(p^2)\right]\gamma^{\mu}}$
as its non-perturbative structure. Only by calculating the vertex in
an arbitrary gauge does this result become clearer. Ball and Chiu have
performed this ${O(\alpha)}$ calculation of the vertex in the Feynman
gauge and we will be able to check their result and correct a couple of
minor misprints in their published work.

    Thus our aim is to compute the fermion-boson vertex to one loop in
perturbation theory in any covariant gauge and to decompose it into its
12 spin components, of these all but 11 are non-zero. This full vertex
is by its very nature free of kinematic singularities. We then divide
the vertex into two parts: the longitudinal and transverse pieces. The
longitudinal component alone fulfills the Ward-Takahashi and Ward identities.
 The way to ensure this without introducing kinematic singularities was
fully described by Ball and Chiu. We then investigate the transverse part
and decompose it into the basis of 8 vectors proposed by
 Ball and Chiu~\cite{BC}.
We examine each coefficient of these and find that two have singularities
in arbitrary gauges. These are not present in the Feynman gauge in
which Ball and Chiu work. We propose a straightforward modification of
their basis that ensures each transverse component is separately free of
kinematic singularities in any covariant gauge. This makes this basis
a natural one for future non-perturbative studies.\\
    We divide the discussion into 5 parts:
\begin{itemize}
\item

the one loop calculation of the vertex, its decomposition into
spin amplitudes and the expression of these in terms of known functions,
including one Spence function with 10 different arguments are all
presented in Sect.~2;

\item

 the one loop calculation of the fermion propagator to determine the
functions ${F(p^2)}$, ${M(p^2)}$, which fix the ${O(\alpha)}$ longitudinal
part of the vertex is in Sect.~3.1;

\item

the extraction of the transverse part of the one loop vertex and its
decomposition into 8 independent components in the Ball-Chiu basis are
described in the rest of Sect.~3;

\item

checking the singularity structure of each of the components of the vertex is
given in Sect.~4.
This leads to the proposal of a new basis for the transverse vertex, which
has coefficients that have only the singularities of the full vertex;

\item

deducing the form of the vertex in specific kinematic regimes.
\end{itemize}
In Sect.~5 we give our brief conclusions.
\section{Perturbative Calculation}
\subsection{Definitions: Feynman rules and basis vectors}
For the most part the definitions given here are standard, but they are stated
here to make this paper self contained. The perturbative calculation
involves the use of bare quantities defined as follows in Minkowski space:\\
\begin{eqnarray}
\mbox{bare vertex:}  -{\it i}e\Gamma^0_{\mu}&=&-{\it i}e\gamma^0_{\mu}\\
\mbox{fermion propagator:} {\it i}{\it S}^0_{F}(p)&=&{\it i}
({\not p+m})/(p^2-m^2)\\
\mbox{photon propagator:} {\it i}{\it \Delta}^0_{\mu\nu}(p)&=&-{\it i}\left[p^2
g_{\mu\nu}+(\xi-1)p_{\mu}p_{\nu}\right]/p^4
\end{eqnarray}
where $e$ is the usual QED coupling and the parameter ${\xi}$ specifies the
covariant gauge.
\vspace{5mm}
\epsfbox[50 50 100 230]{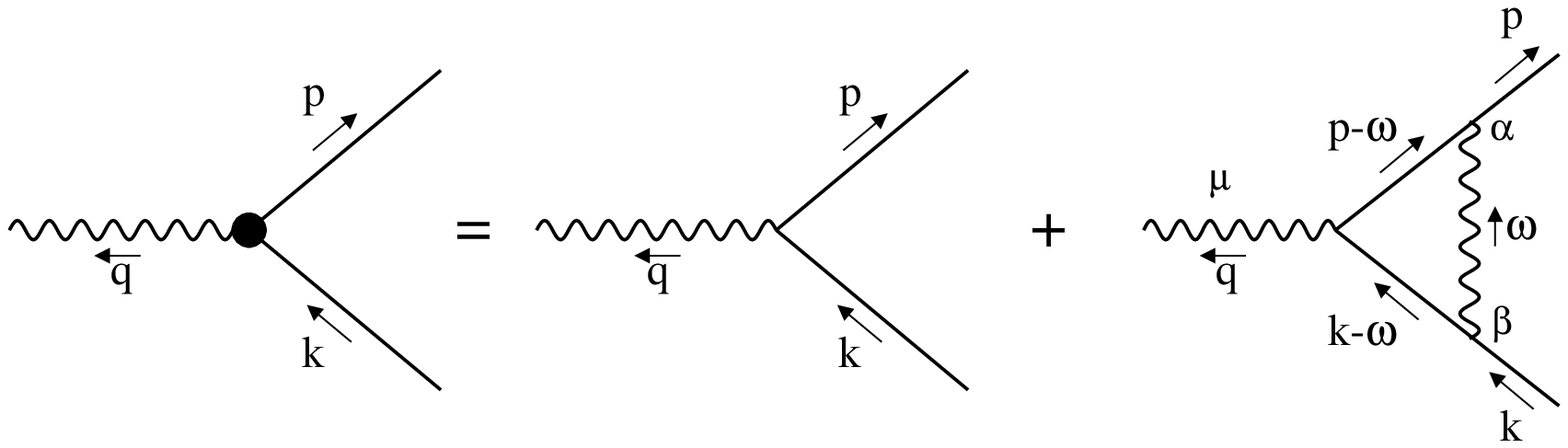}\\
\begin{tabbing}
{\bf{Fig.~1}}: The fermion-boson \=vertex to one loop order showing the
definition of \\
\>momenta\=~ and Lorentz indices.\\
\end{tabbing}

   The vertex, Fig.~1, ${\Gamma^{\mu}(k,p)}$ can be expressed in terms of
12 spin amplitudes formed from the vectors ${\gamma^{\mu},k^{\mu},p^{\mu}}$
and the spin scalars 1,${\not\!k,\not\!p}$ and ${\not\!k\not\!p}$. Thus we can
write
\begin{eqnarray}
\Gamma^{\mu}=\sum^{12}_{i=1} P^{i}V^{\mu}_{i}
\end{eqnarray}
where we number the ${V^{\mu}_{\it i}}$ as follows
\begin{eqnarray}
V_{1}^{\mu}&=&{k^{\mu}}{\not \! k}\;\,\,,\,\,\,\,\,\,
V_{2}^{\mu}={p^{\mu}}{\not \! p}\;\;\,,\,\,\,\,\,\,
V_{3}^{\mu}={k^{\mu}}{\not \! p}\,,\,\,\,\,\,\,
V_{4}^{\mu}={p^{\mu}}{\not \! k}\nonumber\\
V_{5}^{\mu}&=&{\gamma^{\mu}}{\not \! k}{\not \! p},\,\,\,\,\,\,
V_{6}^{\mu}={\gamma^{\mu}}\;\;\;\;\,,\,\,\,\,\,\,
V_{7}^{\mu}={k^{\mu}}\,\,\,\;,\,\,\,\,\,\,
V_{8}^{\mu}={p^{\mu}}\nonumber\\
V_{9}^{\mu}&=&{p^{\mu}}{\not \! k}{\not \! p},\,\,\,\,\,\,
V_{10}^{\mu}={k^{\mu}}{\not \! k}{\not \! p},\,\,\,\,\,\,
V_{11}^{\mu}={\gamma^{\mu}}{\not \! k}\,,\,\,\,\,\,\,
V_{12}^{\mu}={\gamma^{\mu}}{\not \! p}\qquad.
\end{eqnarray}
The vertex satisfies the Ward-Takahashi identity
\begin{eqnarray}
q_{\mu}\Gamma^{\mu}(k,p)={\it S}^{-1}_{F}(k)-{\it S}^{-1}_{F}(p),
\end{eqnarray}
where ${q=k-p}$, and the Ward identity
\begin{eqnarray}
\Gamma^{\mu}(p,p)={\frac{\partial}{\partial p^{\mu}}}{\it S}^{-1}_{F}(p)
\end{eqnarray}
as the non-singular ${k \rightarrow p}$ limit of Eq.~(7). With the fermion
propagator given to any order by Eq.~(1), we follow Ball and Chiu and define
the longitudinal component of the vertex by
\begin{eqnarray}
\Gamma^{\mu}_{L}&=&\frac{\gamma^{\mu}}{2}
\left(\frac{1}{F(k^2)}+\frac{1}{F(p^2)}\right)\nonumber\\
&+&\frac{1}{2}\frac{({\not \! p}+{\not \! k})(k+p)^{\mu}}
{(k^2-p^2)}\left(\frac{1}{F(k^2)}-\frac{1}{F(p^2)}\right)\nonumber\\
&-&\frac{(p+k)^{\mu}}{(k^2-p^2)}\left(\frac{M(k^2)}{F(k^2)}-
\frac{M(p^2)}{F(p^2)}\right)
\end{eqnarray}
${\Gamma^{\mu}_{L}}$ alone then satisfies the Ward-Takahashi identity,
Eq.~(7) and being free of kinematic singularities the Ward identity,
Eq.(8), too.

   The full vertex can then be written as
\begin{eqnarray}
\Gamma^{\mu}(k,p)=\Gamma^{\mu}_{T}(k,p)+\Gamma^{\mu}_{L}(k,p)
\end{eqnarray}
where the transverse part satisfies
\begin{eqnarray}
q_{\mu}\Gamma^{\mu}_{T}(k,p)=0\;\;\;\;\;\mbox{and} \;\;\;\;
\Gamma^{\mu}_{T}(p,p)=0\qquad.
\end{eqnarray}
The Ward-Takahashi identity fixes 4 coefficients of the 12 spin amplitudes
in terms of the fermion functions ---  the 3 combinations explicitly
given in Eq.~(9), while the coefficient of ${\sigma_{\mu\nu}k^{\mu}p^{\nu}}$
must be zero~\cite{BC}. The transverse component ${\Gamma^{\mu}_{T}(k,p)}$
thus involves 8 vectors, which can be expressed in Ball-Chiu form
\footnote{
Ball and Chiu~\cite{BC} use a different notation for the momenta in Fig.~1.
They have
${p}$, ${p^{'}}$ as the incoming and outgoing fermion momenta and ${q}$
as the incoming boson momentum. They define their inverse fermion
propagator ${S_F^{-1}(p)=\rlap/p F(p^2)+G(p^2)}$ and what we here call the
${\tau_i}$ they denote by ${A_i}$.} :
\begin{eqnarray}
\Gamma^{\mu}_{T}(k,p)=\sum^{8}_{i=1} \tau^{i}(k^2,p^2,q^2)T^{\mu}_{i}(k,p)
\end{eqnarray}
where
\begin{eqnarray}
&T^{\mu}_{1}&=p^{\mu}(k\cdot q)-k^{\mu}(p\cdot q)\nonumber\\
&T^{\mu}_{2}&=\left[p^{\mu}(k\cdot q)-k^{\mu}(p\cdot q)\right]({\not\! k}
+{\not\! p})\nonumber\\
&T^{\mu}_{3}&=q^2\gamma^{\mu}-q^{\mu}{\not \! q}\nonumber\\
&T^{\mu}_{4}&=\left[p^{\mu}(k\cdot q)-k^{\mu}(p\cdot q)\right]k^{\lambda}
p^{\nu}{\sigma_{\lambda\nu}}\nonumber\\
&T^{\mu}_{5}&=q_{\nu}{\sigma^{\nu\mu}}\\
&T^{\mu}_{6}&=\gamma^{\mu}(p^2-k^2)+(p+k)^{\mu}{\not \! q}\nonumber\\
&T^{\mu}_{7}&=\frac{1}{2}(p^2-k^2)\left[\gamma^{\mu}({\not \! p}+{\not \! k})
-p^{\mu}-k^{\mu}\right]
+\left(k+p\right)^{\mu}k^{\lambda}p^{\nu}\sigma_{\lambda\nu}\nonumber\\
&T^{\mu}_{8}&=-\gamma^{\mu}k^{\nu}p^{\lambda}{\sigma_{\nu\lambda}}
+k^{\mu}{\not \! p}-p^{\mu}{\not \! k}\nonumber\\
\mbox{with}\;\;\;\;\;\;\;\;\;
&\sigma_{\mu\nu}&=\frac{1}{2}[\gamma_{\mu},\gamma_{\nu}]\qquad.
\end{eqnarray}
The coefficients ${\tau_{\it i}}$ are Lorentz scalar functions of ${k}$ and
${p}$, i.e. functions of ${k^2,p^2,q^2}$.

   A general constraint on the eight ${\tau_i}$'s comes from C-parity
transformations. The full vertex must transform under charge conjugation, C,
in the same way as the bare vertex~\cite{Currents,Rembiesa}, so that
\begin{eqnarray}
C\Gamma_{\mu}(k,p)C^{-1}=-\Gamma_{\mu}^T(-p,-k)\qquad.
\end{eqnarray}
 From the Ward-Takahashi identity, Eq.~(7), it is clear that
${\Gamma_L^{\mu}(k,p)}$ must be symmetric under ${k \leftrightarrow p}$
interchange. The symmetry of the transverse part depends on its
${\gamma}$-matrix structure. Thus Eq.~(15) together with
\begin{eqnarray}
C\gamma_{\mu}C^{-1}=-\gamma_{\mu}^T
\end{eqnarray}
we have from Eqs.~(12,13):
\begin{eqnarray}
\tau_i(k^2,p^2,q^2)&=&\tau_i(p^2,k^2,q^2)\qquad \mbox{for}\qquad
i=1,2,3,4,5,7,8\nonumber\\
\tau_6(k^2,p^2,q^2)&=&-\tau_6(p^2,k^2,q^2)\qquad.
\end{eqnarray}
\subsection{The one loop calculation}

   The vertex of Fig.~1 is naturally expressed as
\begin{eqnarray}
\Gamma^{\mu}(k,p)=\,\gamma^{\mu}+\,\Lambda^{\mu}(k,p).
\end{eqnarray}
Analogous to Eq.~(5) we will express ${\Lambda^{\mu}}$ as
\begin{eqnarray}
\Lambda^{\mu}(k,p)=\sum^{12}_{i=1}P_1^i\,V^{\mu}_i
\label{eq:vertex}
\end{eqnarray}
where the subscript on the ${P^i}$ indicates this calculation is only
to first order in ${\alpha}$.

      From the Feynman rules specified in Sect.~2.1, ${\Lambda^{\mu}}$ to
${O(\alpha)}$ is simply given by:
\begin{eqnarray}
-{\it i}e\Lambda^{\mu}\,=\,\int_{M}\frac{d^4w}{(2\,\pi)^4}
(-{\it i}e\gamma^{\alpha}){\it i}{\it S}^{0}_{F}(p-w)(-{\it i}e\gamma^{\mu})
{\it i}{\it S}^{0}_{F}(k-w)(-{\it i}e\gamma^{\beta}){\it i}
\Delta^0_{\alpha\beta}(w)
\end{eqnarray}
where M denotes the loop integral is to be performed in Minkowski space.
Substituting Eqs.~(3,4) for ${{\it S}^0_{F}(p)}$ and ${\Delta^0_{\mu\nu}(p)}$,
we have with ${\alpha \equiv e^2/4\pi}$:
\begin{eqnarray}
\Lambda^{\mu}&=&-\frac{{\it i}\,{e^2}}{16\,{\pi}^4}\int_{M}d^4w\,
{\gamma}^{\alpha}
\,\frac{(\not \! p-\not\! w+m)}{\left[(p-w)^2-m^2\right]}
\,\gamma^{\mu}\,\frac{(\not \! k-\not\! w+m)}{\left[(k-w)^2-m^2\right]}
\,\gamma^{\beta}\,
\left[\frac{g_{\alpha\beta}}{w^2}+(\xi-1)\frac{w_{\alpha}w_{\beta}}{w^4}\right]
\nonumber\\\\
\Lambda^{\mu}&=&-\frac{{\it i}\,{\alpha}}{4\,{\pi}^3}\int_{M}d^4w\,
\frac{\gamma^{\alpha}\,\left(\,\not \! p-\,\not \! w+\,m\right)\,
\gamma^{\mu}\,\left(\,\not \! k-\,\not \! w+\,m\right)\,\gamma_{\alpha}}
{w^2\,\left[(p-w)^2-m^2\right]\,\left[(k-w)^2-m^2\right]}\nonumber\\
&-&\frac{{\it i}\,{\alpha}}{4\,{\pi}^3}(\xi-1)\int_{M}d^4w\,
\frac{\not \! w\,\left(\,\not \! p-\,\not \! w+\,m\right)\,
\gamma^{\mu}\,\left(\,\not \! k-\,\not \! w+\,m\right)\,\not \! w}
{w^4\,\left[(p-w)^2-m^2\right]\,\left[(k-w)^2-m^2\right]}
\end{eqnarray}
on separating the ${g_{\alpha\beta}}$ and ${w_{\alpha}w_{\beta}}$
parts of the photon propagator.

     What makes the present calculation in an arbitrary covariant gauge
significantly longer and more complicated than that of Ball and Chiu
in the Feynman gauge (${\xi=1}$) is the form of the photon propagator
Eq.~(4). The decomposition of the loop integrals of Eqs.~(20-22) into
 scalar forms in the general case brings greater complexity because
of the potential appearance of infrared divergences in Eq.~(22).
Nevertheless, having the Feynman gauge calculation is a most helpful check
on our results.

    Our first step is to perform a little ${\gamma}$ matrix algebra and
rewrite Eq.~(22) as
\begin{eqnarray}
\Lambda^{\mu}\,&=&\,-\frac{{\it i}\,{\alpha}}{4\,{\pi}^3}\int_{M}d^4w\,
\Bigg\{
\frac{A^{\mu}}{w^2\,\left[(p-w)^2-m^2\right]\,\left[(k-w)^2-m^2\right]}\nonumber\\
&+&(\xi-1)\frac{B^{\mu}}{w^4\,\left[(p-w)^2-m^2\right]\,
\left[(k-w)^2-m^2\right]}
\Bigg\}
\end{eqnarray}
where
\begin{eqnarray}
A^{\mu}\,&=&\,\gamma^{\alpha}\,\left(\,\not \! p-\,\not \! w\right)\,
\gamma^{\mu}\,\left(\,\not \! k-\,\not \! w\right)\,\gamma_{\alpha}\nonumber\\
&+&\,m\,\gamma^{\alpha}\left[({\not \! p}-\,{\not \! w})\,\gamma^{\mu}
+\,\gamma^{\mu}({\not \! k}-\,{\not \! w})\right]\gamma_{\alpha}
+m^2\,\gamma^{\alpha}\,\gamma^{\mu}\,\gamma_{\alpha}
\end{eqnarray}
\begin{eqnarray}
B^{\mu}\,&=&\,\not \! w\,\left(\,\not \! p-\,\not \! w\right)\,
\gamma^{\mu}\,\left(\,\not \! k-\,\not \! w\right)\,\not \! w\nonumber\\
&+&\,m\,\not \! w\left[(\,\not \! p-\,\not \! w)\,\gamma^{\mu}
+\,\gamma^{\mu}(\,\not \! k-\,\not \! w)\right]\not \! w
+\,m^2\not \! w\,\gamma^{\mu}\not \! w\qquad.
\end{eqnarray}
To proceed we introduce the following seven basic integrals over the
loop momentum ${d^{4}w}$~:
${J^{(0)}, J^{(1)}_{\mu}, J^{(2)}_{\mu\nu}, I_0, I^{(1)}_{\mu},
I^{(2)}_{\mu\nu}}$ and ${K^{(0)}}$.
\begin{eqnarray}
{\it J}^{(0)}&=&\int_{M}\,d^4w\,\frac{1}
{w^2\,\left[(p-w)^2-m^2\right]\,\left[(k-w)^2-m^2\right]}\\
{\it J}^{(1)}_{\mu}&=&\int_{M}\,d^4w\,\frac{w_{\mu}}
{w^2\,\left[(p-w)^2-m^2\right]\,\left[(k-w)^2-m^2\right]}\\
{\it J}^{(2)}_{\mu\nu}&=&\int_{M}\,d^4w\,\frac{w_{\mu}w_{\nu}}
{w^2\,\left[(p-w)^2-m^2\right]\,\left[(k-w)^2-m^2\right]}\\
{\it I}^{(0)}&=&\int_{M}\,d^4w\,\frac{1}
{w^4\,\left[(p-w)^2-m^2\right]\,\left[(k-w)^2-m^2\right]}\\
{\it I}^{(1)}_{\mu}&=&\int_{M}\,d^4w\,\frac{w_{\mu}}
{w^4\,\left[(p-w)^2-m^2\right]\,\left[(k-w)^2-m^2\right]}\\
{\it I}^{(2)}_{\mu\nu}&=&\int_{M}\,d^4w\,\frac{w_{\mu}w_{\nu}}
{w^4\,\left[(p-w)^2-m^2\right]\,\left[(k-w)^2-m^2\right]}\\
{\it K}^{(0)}&=&\int_{M}\,d^4w\,\frac{1}
{\left[(p-w)^2-m^2\right]\,\left[(k-w)^2-m^2\right]}
\end{eqnarray}
${\Lambda^{\mu}}$ of Eq.~(22) can then be re-expressed in terms of five
of these as:
\begin{eqnarray}
\Lambda^{\mu}\,=\,&-&\frac{{\it i}\,{\alpha}}{4\,{\pi}^3}
\Bigg\{
\left[\gamma^{\alpha}\left(
{\not \! p}\,{\gamma^{\mu}}\,{\not \! k}
+\,m{\not \! p}\,{\gamma^{\mu}}+\,m{\gamma^{\mu}}\,{\not \! k}
+\,m^2{\gamma^{\mu}}
\right)\gamma_{\alpha}
\right]{\bf {\it J}}^{(0)}\hspace{50mm}\nonumber\\
&& \hspace{10mm}
-\left[{\gamma^{\alpha}}\left(
{\not \! p}\,{\gamma^{\mu}}{\gamma^{\nu}}
+{\gamma^{\nu}}{\gamma^{\mu}}{\not \! k}
+m\,{\gamma^{\nu}}{\gamma^{\mu}}+m{\gamma^{\mu}}{\gamma^{\nu}}\right)
\gamma_{\alpha}\right]{\bf {{\it J}_{\nu}^{(1)}}}\nonumber\\
&& \hspace{10mm}
+{\gamma^{\alpha}}{\gamma^{\nu}}{\gamma^{\mu}}{\gamma^{\lambda}}
\gamma_{\alpha}{\bf {{\it J}_{\nu\lambda}^{(2)}}}\nonumber\\
&& \hspace{10mm}
+(\xi-1)\Bigg[
\left(-{\gamma^{\nu}}{\not \! p}\,{\gamma^{\mu}}-\,
{\gamma^{\mu}}{\not \! k}\,{\gamma^{\nu}}
-m\,{\gamma^{\mu}}{\gamma^{\nu}}-m\,{\gamma^{\nu}}{\gamma^{\mu}}\right)
{\bf {\it J}_{\nu}^{(1)}}
+{\gamma^{\mu}}{\bf \it K}^{(0)}\nonumber\\
&& \hspace{25mm}
+\left({\gamma^{\nu}}{\not \! p}\,{\gamma^{\mu}}{\not \! k}\,{\gamma^{\lambda}}
+m{\gamma^{\nu}}{\not \! p}\,{\gamma^{\mu}}{\gamma^{\lambda}}
+m{\gamma^{\nu}}{\gamma^{\mu}}{\not \! k}\,{\gamma^{\lambda}}
+m^2{\gamma^{\nu}}{\gamma^{\mu}}{\gamma^{\lambda}}\right)
{\bf {{\it I}_{\nu\lambda}^{(2)}}}
\Bigg]
\Bigg\}.
\end{eqnarray}

   Our next step is to compute the basic integrals of Eqs.~(26-32),
   \cite{tHooft,DD,Davy}
each of
which is a function of k and p. We relegate to the Appendix A the tabulation of
each of the intermediate integrals.\\ \\
\vspace{7mm}
{\underline{\bf{${{\it J}^{(1)}_{\mu}}$ and ${{\it J}^{(2)}_{\mu\nu}}$
calculated:}}}

     The method of relating Lorentz vector and tensor integrals to scalar
integrals is by now standard~\cite{BC}, so we will not dwell on this but
merely give one example to serve as a reminder to the reader. ${J^{(1)}_{\mu}}$
of Eq.~(27) can as a Lorentz vector only have components in the directions
of the 4-momenta ${k_{\mu}}$ and ${p_{\mu}}$. Thus, we can write
\begin{eqnarray}
{\it J}^{(1)}_{\mu}=\frac{{\it i}\pi^2}{2}\left[
k_{\mu}J_{A}(k,p)+p_{\mu}J_{B}(k,p)\right]
\end{eqnarray}
where ${J_{A}, J_{B}}$ must be scalar functions of ${k}$ and ${p}$.
The factor of
${{\it i}\pi^2/2}$ is taken out purely for later convenience. It is then
easy to see that
\begin{eqnarray}
J_{A}(k,p)=\frac{1}{{\it i}\pi^2\Delta^2}\left[2\,k\cdot p\, p^{\mu}J^{(1)}_
{\mu}-2\,p^2k^{\mu}J^{(1)}_{\mu}\right]
\end{eqnarray}
with a similar expression for ${J_{B}}$,\\
where
\begin{eqnarray}
\Delta^2=(k\cdot p)^2-k^2p^2
\end{eqnarray}
is the ubiquitous triangle function of ${k}$, ${p}$ and ${q}$. One then
rewrites the integrand numerators using, for instance
\begin{eqnarray}
2p\cdot w\,=\,p^2+w^2-m^2-\left[(p-w)^2-m^2\right]
\end{eqnarray}
so that in ${d}$-dimension
\vspace{5mm}
\begin{eqnarray}
J_{A}(k,p)&=&\frac{1}{{\it i}\pi^2\Delta^2}\Bigg\{
\left[k\cdot p\left(p^2-m^2\right)-p^2\left(k^2-m^2\right)\right]
\int\frac{d^dw}{w^2[(k-w)^2-m^2][(p-w)^2-m^2]}\nonumber\\
&+&\left(k\cdot p-p^2\right)\int\frac{d^dw}{[(k-w)^2-m^2][(p-w)^2-m^2]}
\nonumber\\
&-&k\cdot p\int\frac{d^dw}{w^2[(k-w)^2-m^2]}
\;+\;p^2\int\frac{d^dw}{w^2[(p-w)^2-m^2]}
\Bigg\}\qquad.
\end{eqnarray}
The basic 16 scalar integrals, of which four appear in this equation
${Q_{7}(k,p), Q_{8}(k,p), Q_{14}(k,p) and Q_{14}(p,k)}$, are given in the
Appendix A. We thus deduce
\begin{eqnarray}
J_{A}(k,p)&=&\frac{1}{\Delta^2}\Bigg\{
\frac{{\it J}_{0}}{2}\left(-m^2\,p\cdot q-\,p^2\,k\cdot q\right)
+k\cdot p\,{\it L^{'}}-p^2{\it L}-2\,p\cdot q\,{\it S}
\Bigg\}\\ \nonumber\\
J_{B}(k,p)&=&J_{A}(p,k)
\end{eqnarray}
where
\begin{eqnarray}
&{\it J}^{(0)}&=\frac{{\it i}\,\pi^2}{2}{\it J}_{0}\\
&{\it L}&=\left(1-\frac{m^2}{p^2}\right)\,\ln \left(1-\frac{p^2}{m^2}\right)
\\
&{\it L}^{'}&={\it L}\left(p\leftrightarrow k\right)\\
&{\it S}&=\frac{1}{2}\left(1-4\,\frac{m^2}{q^2}\right)^{1/2}\,
\ln \frac{\left[\left(1-{4m^2}/{q^2}\right)^{1/2}+1\right]}
{\left[\left(1-{4m^2}/{q^2}\right)^{1/2}-1\right]}
\end{eqnarray}
with ${J_0}$ being expressed in terms of Spence
 functions~\cite{tHooft,DD,Davy,Spence} --- see Appendix A, Eqs.~(A15-A18).
In an analogous
fashion, the tensor integral ${J^{(2)}_{\mu\nu}}$ of Eq.~(28) can be expressed
in terms of scalar integrals ${K}$,${J_{C},J_{D}}$ and ${J_{E}}$ by
\begin{eqnarray}
{\it J}^{(2)}_{\mu\nu}&=&\frac{{\it i}\pi^2}{2}\Bigg\{
\frac{g_{\mu\nu}}{d}K_0+\left(k_{\mu}k_{\nu}-g_{\mu\nu}\frac{k^2}{4}\right)
J_{C}\nonumber\\
&+&\left(p_{\mu}k_{\nu}+k_{\mu}p_{\nu}-g_{\mu\nu}\frac{(k\cdot p)}{2}\right)
J_{D}+\left(p_{\mu}p_{\nu}-g_{\mu\nu}\frac{p^2}{4}\right)J_{E}
\Bigg\}\qquad.
\end{eqnarray}
All but $K(k,p)$ are ultraviolet finite and so the number of dimensions ${d}$
has been equal to 4. In ${d\equiv4+\epsilon}$ dimensions, with
${\mu}$ the usual scale parameter introduced to ensure the coupling ${\alpha}$
remains dimensionless for any $d$, we have
\begin{eqnarray}
K_{0}(k,p)&=&\frac{2}{i\pi^2}K^{(0)}\\\nonumber\\
K_{0}(k,p)&=&2\mu^{\epsilon}\left[C-2S+2\right]
\end{eqnarray}
where
\begin{eqnarray}
C=-\frac{2}{\large\epsilon }-\gamma-\ln(\pi)-\ln({m^2}/{\mu^2}).
\end{eqnarray}
Then
\begin{eqnarray}
J_{C}(k,p)&=&\frac{1}{4\,\Delta^2}\Bigg\{
\left(2\,p^2+2\,k\cdot p\,\frac{m^2}{k^2}\right)-4\,k\cdot p\,{\it S}
+2\,k\cdot p\,\left(1-\frac{m^2}{k^2}\right){\it L^{'}}\nonumber\\\nonumber \\
&& \hspace{10mm}
+\left(2\,k\cdot p\,(p^2-m^2)+
3\,p^2\,(m^2-k^2)\right)J_{A}+\,p^2(m^2-p^2)J_{B}
\Bigg\},\\ \nonumber\\
J_{D}(k,p)&=&\frac{1}{4\Delta^2}\Bigg\{
2\,k\cdot p\,\left[(k^2-m^2)J_{A}
+(p^2-m^2)J_{B}-\,1\right]\nonumber\\\nonumber \\
&& \hspace{10mm}
-k^2\left[2\,\frac{m^2}{k^2}-2{\it S}
+\left(1-\frac{m^2}{k^2}\right){\it L^{'}}
+\,(p^2-m^2)J_{A}\right]\nonumber\\ \nonumber\\
&& \hspace{10mm}
-p^2\left[-2\,{\it S}
+\left(1-\frac{m^2}{p^2}\right){\it L}
+\,(k^2-m^2)J_{B}\right]
\Bigg\},\\ \nonumber\\
J_{E}(k,p)&=&J_{C}(p,k)
\end{eqnarray}
all of which involve the previously defined ${J_{A},J_{B},L,L^{'}}$ and ${S}$
of Eqs.~(42-44).\\\\
\vspace{7mm}
{\underline{\bf{${{\it I}^{(1)}_{\mu}}$ and ${{\it I}^{(2)}_{\mu\nu}}$
calculated:}}}\\
\indent
In a way analogous to the computation of ${J^{(1)}_{\mu}}$ and
${J^{(2)}_{\mu\nu}}$ the ultraviolet finite integrals ${I^{(1)}_{\mu}}$ and
${I^{(2)}_{\mu\nu}}$~\cite{Davy} of Eqs.~(30,31) can be re-expressed in terms
of
scalar
integrals, ${I_{A}, I_{B}, I_{C}, I_{D}, I_{E}}$, that in turn involve the
same functions we have already computed. Thus
\begin{eqnarray}
{\it I}^{(1)}_{\mu}&=&\frac{{\it i}\pi^2}{2}\left[
 k_{\mu}I_{A}(k,p)+p_{\mu}I_{B}(k,p)\right]
\end{eqnarray}
where
\begin{eqnarray}
I_{A}(k,p)&=&\frac{1}{\Delta^2}\Bigg\{
-\frac{k\cdot q}{2}{{\it J}_{0}}
-\frac{2q^2}{\chi}\left\{(m^2-p^2)k^2-(m^2-k^2)\,k\cdot p\right\}{\it S}
\nonumber\\
&& \hspace{10mm}
+\frac{1}{(m^2-p^2)}\left[p^2-k\cdot p
+\frac{p^2\,q^2}{\chi}(k^2-m^2)\left(m^2+k\cdot p\right)\right]{\it L}
\nonumber\\
&& \hspace{10mm}
+\frac{k^2\,q^2}{\chi}\left(m^2+k\cdot p\right){\it L^{'}}
\Bigg\}\nonumber\\
\end{eqnarray}
and
\begin{eqnarray}
I_{B}(k,p)&=&I_{A}(p,k)\hspace{97mm}
\end{eqnarray}
with the denominator
\begin{eqnarray}
\chi&=&(q^2-2m^2)(p^2-m^2)(k^2-m^2)+m^2(p^2-m^2)^2+m^2(k^2-m^2)^2\nonumber\\
&=&p^2k^2q^2+2\left[(p^2+k^2)\,k\cdot p-2p^2k^2\right]m^2+m^4q^2
\end{eqnarray}
\begin{eqnarray}
{\it I}^{(2)}_{\mu\nu}&=&\frac{{\it i}\pi^2}{2}\Bigg\{
\frac{g_{\mu\nu}}{4}{\it J}_{0}+\left(k_{\mu}k_{\nu}
-g_{\mu\nu}\frac{k^2}{4}\right)I_{C}
\nonumber\\
&& \hspace{10mm}
+\left(p_{\mu}k_{\nu}+k_{\mu}p_{\nu}-g_{\mu\nu}\frac{(k\cdot p)}{2}\right)I_{D}
+\left(p_{\mu}p_{\nu}-g_{\mu\nu}\frac{p^2}{4}\right)I_{E}
\Bigg\}\\ \nonumber \\
I_{C}(k,p)&=&\frac{1}{4\Delta^2}\Bigg\{
2p^2{\it J}_{0}-4\,\frac{k\cdot p}{k^2}\left(1+\frac{m^2}{(k^2-m^2)}
{\it L^{'}}\right)\nonumber\\
&& \hspace{10mm}
+\left\{2\,k\cdot p-3p^2\right\}J_{A}-p^2J_{B}\nonumber\\
&& \hspace{10mm}
+\left\{-2\,k\cdot p\,(m^2-p^2)+3p^2(m^2-k^2)\right\}I_{A}+p^2(m^2-p^2)I_{B}
\Bigg\}\\ \nonumber\\
I_{D}(k,p)&=&\frac{1}{4\,\Delta^2}\Bigg\{
-2(k\cdot p){\it J}_{0}+2\left(1+\frac{m^2}{(k^2-m^2)}{\it L^{'}}\right)
+2\left(1+\frac{m^2}{(p^2-m^2)}{\it L}\right)\nonumber\\ \nonumber\\
&& \hspace{10mm}
+\left(2\,k\cdot p-k^2\right)J_{A}+\,\left(2\,k\cdot p-p^2\right)J_{B}
\nonumber\\
&& \hspace{10mm}
+\left[k^2(m^2-p^2)-2\,k\cdot p\,(m^2-k^2)\right]I_{A}\nonumber\\
&& \hspace{10mm}
+\left[p^2(m^2-k^2)-2\,k\cdot p\,(m^2-p^2)\right]I_{B}
\Bigg\}\\
I_{E}(k,p)&=&I_{C}(p,k).
\end{eqnarray}
\noindent The $1/\chi$ term in $I_A, I_B, I_c$ and $I_D$ arise from the extra
$1/w^2$ factor that occurs in the second integral of Eq.~(23).  Notice that the
$1/\chi$ term arises in all but the Feynman gauge.  The possibility of
singularities
 at $\chi = 0$ has consequences as we shall see later.
\newpage
\noindent
{\underline{\bf{${\Lambda^{\mu}}$ collected:}}}\\
\indent
In terms of the basic functions ${{\it J}_0, {\it J}_A, {\it J}_B, {\it J}_C,
 {\it J}_D, {\it J}_E, {\it I}_0, {\it I}_A, {\it I}_B, {\it I}_C,{\it I}_D,
 {\it I}_E}$ and the ultraviolet divergent ${K_0}$, all of which depend on the
momenta $k$ and $p$, i.e. are functions of the Lorentz scalars ${k^2, p^2}$ and
${q^2, \Lambda^{\mu}}$ can be written completely with its gamma-matrix
and Lorentz index structure displayed explicitly~:
\begin{eqnarray}
\Lambda^{\mu}(k,p)&=&\sum^{12}_{i=1}P_1^i\,V^{\mu}_i
\hspace{100mm}
\mbox{{\protect(\ref{eq:vertex}})}\nonumber\\\nonumber\\
P_1^1&=&2J_{A}-2J_{C}+(\xi-1)\left(m^2I_{C}+p^2I_{D}\right)\nonumber\\
P_1^2&=&2J_{B}-2J_{E}+(\xi-1)\left(k^2I_{D}+m^2I_{E}\right)\nonumber\\
P_1^3&=&-2{\it J}_{0}+2J_{A}+2J_{B}-2J_{D}\nonumber\\
&+&(\xi-1)\left(-\frac{{\it J}_0}{2}-k^2I_{C}
+\frac{k^2}{2}I_{C}
-\,k\cdot p\,I_{D}+m^2I_{D}+\frac{p^2}{2}I_{E}+J_{A}\right)\nonumber\\
P_1^4&=&-2J_{D}+(\xi-1)\left(k^2I_{C}+m^2I_{D}-J_{A}\right)\nonumber\\
P_1^5&=&J_{0}-J_{A}-J_{B}\nonumber
+(\xi-1)\left(\frac{{\it J}_0}{4}
+\frac{k^2}{4}I_{C}
+\frac{k\cdot p}{2}I_{D}+\frac{p^2}{4}I_{E}-\frac{1}{2}J_{A}
-\frac{1}{2}J_{B}\right)\nonumber\\
P_1^6&=&\left(-m^2J_{0}-k^2J_{A}-p^2J_{B}+\frac{k^2}{2}J_{C}
+k\cdot
p\,J_{D}+\frac{p^2}{2}J_{E}+2\left(1+\epsilon\right)K_0\right)\nonumber\\
&+&(\xi-1)\Bigg(-m^2\frac{{\it J}_0}{2}-\frac{m^2}{4}k^2I_{C}
-\frac{m^2}{2}\,k\cdot p\,I_{D}\nonumber\\
&-&\frac{m^2}{4}p^2I_{E}-\frac{k^2}{2}J_{A}-\frac{p^2}{2}J_{B}
+\left[C+2-2{\it S}\right]\Bigg)\nonumber\\
P_1^7&=&2mJ_{0}-4mJ_{A}\nonumber\\
&+&(\xi-1)m\Bigg(2\frac{{\it J}_0}{2}-2\,k\cdot p\,I_{C}
+\frac{k^2}{2}I_{C}-p^2I_{D}\nonumber\\
&-&\,k\cdot p\,I_{D}-\frac{p^2}{2}I_{E}-J_{A}\Bigg)\nonumber\\
P_1^8&=&2mJ_{0}-4mJ_{B}\nonumber\\
&+&(\xi-1)m\left(2\frac{{\it J}_0}{2}+\frac{k^2}{2}I_{C}
-\,k\cdot p\,I_{D}+k^2I_{D}
-\frac{p^2}{2}I_{E}-J_{B}\right)\nonumber\\\nonumber\\
P_1^9&=&(\xi-1)m\left(I_{D}+I_{E}\right)\nonumber\\
P_1^{10}&=&(\xi-1)m\left(I_{D}+I_{C}\right)\nonumber\\
P_1^{11}&=&(\xi-1)m\left(p^2I_{D}+\,k\cdot p\,I_{C}
+\frac{p^2}{2}I_{E}-\frac{k^2}{2}I_{C}
\right)\nonumber\\
P_1^{12}&=&(\xi-1)m\left(-k^2I_{D}-\,k\cdot p\,I_{E}
+\frac{p^2}{2}I_{E}-\frac{k^2}{2}I_{C}
\right)\qquad.
\end{eqnarray}
Notice that both the integrals ${I_A, I_B}$ cancel out in this result.
 Though this expression appears to involve all 12 spin vectors, one of their
coefficients is not independent. The Ward-Takahashi identity, Eq.~(7), only
involves ${\not\!k, \not\!p,1}$ as spin structure on the right hand side. This
means that ${\not\!k\not\!p}$ and ${\not\!p\not\!k}$ terms that occur in
${q_{\mu}\Gamma^{\mu}}$ from Eq.~(7) must occur in the form of the
anticommutator ${\left\{\not\!k,\not\!p\right\}=2k\cdot p}$. Consequently;
the coefficients $P_i$ of Eq.~(19) are related by:
\begin{eqnarray}
P_1^{12}=P_1^{9}(p^2-k\cdot p)+P_1^{10}(k\cdot p-k^2)-P_1^{11}
\end{eqnarray}
Formally, this completes our calculation of the one loop corrections to the
QED vertex in any covariant gauge.
\section{Analytic Structure of the Vertex}
\subsection{The Longitudinal vertex}
%
%
\epsfbox[50 50 100 230]{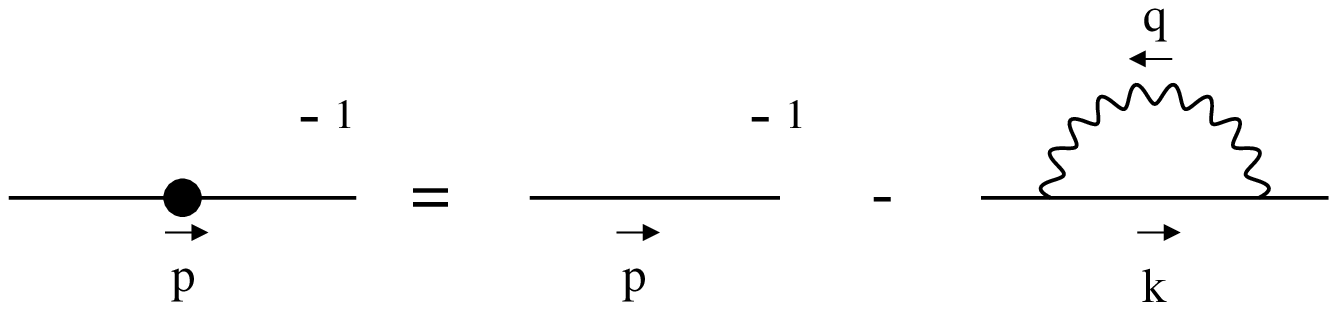}
{\bf{Fig.~2}}: The inverse fermion propagator to one loop order in
perturbation theory.\\ \\
As explained in Sect.~2.1, the longitudinal component of the vertex is
determined by the fermion functions, ${F(p^2), M(p^2)}$, thanks to the
Ward-Takahashi identity. In this section we compute these functions to
${O(\alpha)}$ by calculating the one loop corrections to the fermion
propagator, Fig.~2. Straightforward calculation yields
\begin{eqnarray}
F^{-1}(p^2)&=&1+\frac{\alpha\xi}{4\pi}\left[C\mu^{\epsilon}
+\left(1+\frac{m^2}{p^2}\right)\left(1-L\right)\right]\\
M(p^2)&=&m+\frac{\alpha m}{\pi}\left[\left(1+\frac{\xi}{4}\right)
+\frac{3}{4}(C\mu^{\epsilon}-L)
-\frac{\xi}{4}\frac{m^2}{p^2}(1-L)\right]
\end{eqnarray}
with the same factors $C$ and ${\it L}$ of Eqs.~(48,42). Simple substitution
into Eq.~(9) gives the longitudinal vertex, which we write out as
\begin{eqnarray}
\Gamma^{\mu}_{L}&=&\frac{\alpha\xi}{8\,\pi}{\gamma^{\mu}}\left[
2C{\mu}^{\epsilon}+\left(1+\frac{m^2}{k^2}\right)(1-{\it L^{'}})
+\left(1+\frac{m^2}{p^2}\right)(1-{\it L})\right]\nonumber\\
&+&\frac{\alpha\xi}{4\,\pi}(k^{\mu}{\not \! p}+k^{\mu}{\not \! k}
+p^{\mu}{\not \! p}
+p^{\mu}{\not \! k})\frac{1}{2\,(k^2-p^2)}\left[m^2\left(\frac{1}{k^2}
-\frac{1}{p^2}\right)-\left(1+\frac{m^2}{k^2}\right){\it L^{'}}
+\left(1+\frac{m^2}{p^2}\right){\it L}\right]\nonumber\\
&-&\frac{\alpha\,m}{4\pi}\left(3+\xi\right)
\frac{(p+k)^{\mu}}{(k^2-p^2)}\left[\it L-{\it L^{'}}
\right]\\
\end{eqnarray}
\subsection{The Transverse Vertex}
Having calculated the vertex ${O(\alpha)}$ , Eq.~(19), we can subtract from
it the longitudinal vertex of Sect.~3.1, Eq.~(64) and obtain (Eq.~(12)) the
transverse vertex to ${O(\alpha)}$ . This is given by a rather
lengthy expression
\begin{eqnarray}
\Gamma^{\mu}_{T}(k,p)&=&\frac{\alpha}{4\,\pi}\Bigg\{
\sum_{{\it i}=1}^{12}\,V^{\mu}_{\it i}\Bigg(
\frac{1}{2\,\Delta^2}
\left[a^{(\it i)}_{1}+(\xi-1)a^{(\it i)}_{2}\right]J_{A}\nonumber\\
&& \hspace{23mm}
+\frac{1}{2\,\Delta^2}
\left[b^{(\it i)}_{1}+(\xi-1)b^{(\it i)}_{2}\right]J_{B}\nonumber\\
&& \hspace{23mm}
+\frac{1}{2\,\Delta^2}
\left[c^{(\it i)}_{1}+(\xi-1)c^{(\it i)}_{2}\right]I_{A}\nonumber\\
&& \hspace{23mm}
+\frac{1}{2\,\Delta^2}
\left[d^{(\it i)}_{1}+(\xi-1)d^{(\it i)}_{2}\right]I_{B}\nonumber\\
&& \hspace{23mm}
+\frac{1}{2\,p^2\,(k^2-p^2)\,(p^2-m^2)\,\Delta^2}
\left[e^{(\it i)}_{1}+(\xi-1)e^{(\it i)}_{2}\right]{\it L}\nonumber\\
&& \hspace{23mm}
+\frac{1}{2\,k^2\,(k^2-p^2)\,(k^2-m^2)\,\Delta^2}
\left[f^{(\it i)}_{1}+(\xi-1)f^{(\it i)}_{2}\right]{\it L^{'}}\nonumber\\
&& \hspace{23mm}
+\frac{1}{\Delta^2}
\left[g^{(\it i)}_{1}+(\xi-1)g^{(\it i)}_{2}\right]{\it S}\nonumber\\
&& \hspace{23mm}
+\frac{1}{2\Delta^2}
\left[h^{(\it i)}_{1}+(\xi-1)h^{(\it i)}_{2}\right]{\it J}_{0}\nonumber\\
&& \hspace{23mm}
+\frac{1}{\,\Delta^2}
\left[l^{(\it i)}_{1}+(\xi-1)l^{(\it i)}_{2}\right]
\Bigg)
\Bigg\}
\end{eqnarray}
in terms of the 12 vectors ${V^{\mu}_{{\it i}}}$ of Eq.~(5) with the
coefficients which are listed in the Appendix B.
\newpage
Our task is then to express this result in terms of the 8 basis vectors
defining ${\Gamma^{\mu}_{T}(k,p)}$, Eq.~(12). Thus
from Eq.~(13) we can alternatively write out
\begin{eqnarray}
\Gamma^{\mu}_{T}&=&k^{\mu}{\not \! k}\left[\tau_{2}(p^2-k\cdot p)-\tau_{3}
+\tau_{6}\right]\nonumber\\
&+&p^{\mu}{\not \! p}\left[\tau_{2}(k^2-k\cdot p)-\tau_{3}-\tau_{6}\right]
\nonumber\\
&+&k^{\mu}{\not \! p}\left[\tau_{2}(p^2-k\cdot p)+\tau_{3}-\tau_{6}
+\tau_{8}\right]\nonumber\\
&+&p^{\mu}{\not \! k}\left[\tau_{2}(k^2-k\cdot p)+\tau_{3}
+\tau_{6}-\tau_{8}\right]\nonumber\\
&+&\gamma^{\mu}\left[\tau_{3}q^2+\tau_{6}(p^2-k^2)+\tau_{8}(k\cdot p)\right]
\nonumber\\
&+&\gamma^{\mu}{\not \! k}{\not \! p}\left[-\tau_{8}\right]\nonumber\\
&+&p^{\mu}\left[\tau_{1}(k^2-k\cdot p)-\tau_{4}(k^2-k\cdot p)(k\cdot
p)-\tau_{5}
+\frac{\tau_{7}}{2}(k^2-p^2-2k\cdot p)\right]\nonumber\\
&+&k^{\mu}\left[\tau_{1}(p^2-k\cdot p)-\tau_{4}(p^2-k\cdot p)(k\cdot
p)+\tau_{5}
+\frac{\tau_{7}}{2}(k^2-p^2-2k\cdot p)\right]\nonumber\\
&+&p^{\mu}{\not \! k}{\not \! p}\left[\tau_{4}(k^2-k\cdot p)+\tau_{7}\right]
\nonumber\\
&+&k^{\mu}{\not \! k}{\not \! p}\left[\tau_{4}(p^2-k\cdot p)+\tau_{7}\right]
\nonumber\\
&+&\gamma^{\mu}{\not \! k}\left[-\tau_{5}+\frac{\tau_{7}}{2}(p^2-k^2)\right]
\nonumber\\
&+&\gamma^{\mu}{\not \! p}\left[\tau_{5}+\frac{\tau_{7}}{2}(p^2-k^2)\right]
\qquad.
\end{eqnarray}
Comparing Eqs.~(5) and (65), we have 12 equations for the 8 unknown
${\tau_{i}}$. Since ${\Gamma^{\mu}_{T}}$ is transverse to the vector
${q_{\mu}}$, Eq.~(11), only 8 of these equations are independent.
Laborious solution yields expressions for the 8 transverse coeffecients
${\tau_{i}}$. Each is a function of ${k^2, p^2, q^2}$ and ${\xi}$. The
results are as follows:
\begin{eqnarray}
\tau_{1}=\frac{\alpha}{4\pi}\frac{\left (3+\xi\right)}{\Delta^2}m\Bigg\{
&-&\frac{1}{2}\left[m^{2}+k\cdot p\right]{\it J}_{0}
-{2{\it S}} \nonumber\\ \nonumber\\
&+&\frac {\left [p^{2}+\,k\cdot p\,\right]{\it L}}{\left (p^{2}-k^{2}\right )}
-\frac {\left[k^{2}+\,k\cdot p\,\right]{\it L^{'}}}{\left(p^{2}-k^{2}\right )}
\Bigg\}\qquad , \hspace{40mm}
\end{eqnarray}
\newpage
\begin{eqnarray}
\tau_{2}=&-&\frac{3}{4\Delta^2} \left[\,m^{2}+\,k\cdot p\right]
\tau_{8} \nonumber\\
&+&\frac{\alpha}{4\pi\Delta^{2}}\Bigg\{
\frac {1}{8}\left(q^{2}-4m^{2}\right){\it J}_{0}+\frac{({\it L}+{\it
L^{'}})}{4}-\frac{1}{2}-\frac {m^2}{\,2\,p^{2}k^{2}}\,k\cdot p \nonumber\\
&& \hspace{10mm} +\frac{1}{2\,(p^2-k^2)}\left[
[k^2+\,k\cdot p\,]\,\left(1+\,\frac{m^2}{k^2}\right){\it L^{'}}
-\,[p^2+\,k\cdot p\,]\left(1+\frac{m^2}{p^2}\right){\it L}
\right]\Bigg\} \nonumber\\ \nonumber\\
&+&\frac{\alpha}{8\pi\Delta^2}\left(\xi-1\right)\Bigg\{
%
%
\left[\frac{1}{2}\left(p^2+\,k^2\right)\,+\,\frac{3q^2}{4\Delta^{2}}\,\left(p^2k^2-\,m^4\right)\right]{\it J}_{0}\,+\,1\,-\,\frac{m^2\,}{p^2\,k^2}\,k\cdot p
\nonumber\\ \nonumber   \\
&& \hspace{25mm} +\frac{1}{\chi} \Bigg[
\frac{-2m^2}{(p^2-k^2)}
\left\{\left (p^2+\,m^2\,\frac{k^2}{p^2}\right ){\it L}
\,-\,\left (k^2+\,m^2\,\frac{p^2}{k^2}\right ){\it L^{'}}\right\}\,\Delta^{2}
\nonumber \\ \nonumber \\
&& \hspace{25mm} -q^2\,\frac{[p^2+2\,k\cdot p\,+k^2]}{2\,(p^2-k^2)}
\left \{m^6\left (\frac{{\it L}}{p^2}-\frac{{\it
L^{'}}}{k^2} \right )\,+\,p^2k^2\,({\it L}-{\it L^{'}})\right \} \nonumber\\
\nonumber\\
&& \hspace{25mm}
+\frac{q^2}{2}m^4\left\{\left(1-\frac{m^2}{p^2}\right)
{\it L}+\left(1-\frac{m^2}{k^2}\right){\it L^{'}}\right\}
-\frac{m^2}{2}(p^4-k^4)({\it L} -{\it L^{'}})\nonumber\\
&& \hspace{25mm}
+\frac{q^2}{2}\,\left[k^2p^2\left({\it L}+{\it L^{'}}\right)
\,-\,m^2\left(k^2{\it L}\,+\,p^2{\it L^{'}}\right)\right] \nonumber\\
&& \hspace{25mm}
+\frac{3\,m^2}{2}\,(p^2-k^2)\,\left[(m^2-p^2)\,{\it L}
\,-\,(m^2-k^2)\,{\it L^{'}}\right]\nonumber \\
&& \hspace{25mm}
+\,\frac{3q^4}{4\Delta^2}
\,(m^4-p^2k^2)\left[(m^2-p^2){\it L}+(m^2-k^2){\it L^{'}}\right]\nonumber\\
&& \hspace{25mm}
+\frac{3\,q^2}{4\,\Delta^2}\,(p^2-k^2)\,(m^4-p^2\,k^2)\,\left[(m^2+p^2)\,{\it
L}\,-\,(m^2+k^2)\,{\it L^{'}}\right] \nonumber\\
&& \hspace{25mm}
+\Bigg[
-8\,q^2(\,m^4+\,k^2p^2)\,+\,12p^2k^2(\,q^2-\,m^2)+\,2q^2(\,k^2+\,p^2)m^2
\nonumber\\
&& \hspace{25mm}
-\frac{3}{\Delta^2}m^4q^4(m^2+k\cdot p\,)\,+\,
\frac{3}{\Delta^2}k^2p^2q^2\,k\cdot p\,(q^2-m^2)+\,2(p^2-k^2)^2m^2\nonumber\\
&& \hspace{25mm}
-\frac{9}{\Delta^2}m^2p^2k^2q^2\,k\cdot p\,+\,
\frac{3}{\Delta^2}m^2p^2k^2(p^2-k^2)^2\Bigg]{\it S}
\Bigg]
\Bigg\} \qquad ,
\end{eqnarray}
\newpage
\begin{eqnarray}
\tau_{3}=&-& \frac{3}{32m\Delta^2}\,[\,m^2+\,k\cdot p\,]\,(p^2-k^2)^2\,
\tau_{1}(\xi=1)\nonumber\\
&+&\frac{\alpha}{8\pi\Delta^2}\,\Bigg\{
\left[-\Delta^2-\,\frac{(\,p^2+\,k^2-\,2\,m^2)^2}{8}\right]{\it J}_{0}\,
-\,2\,[\,m^2+\,k\cdot p\,]{\it S}\nonumber\\
&& \hspace{15mm}
+\frac{k\cdot p}{2}\left[\left(1-\frac{m^2}{p^2}\right){\it L}\,
+\,\left(1-\frac{m^2}{k^2}\right){\it L^{'}}\right]
+\,\frac{1}{4}(p^2-k^2)\,({\it L}-{\it L^{'}})\nonumber\\
&& \hspace{15mm}
+\,[\,m^2+\,k\cdot p]\,+\,
\frac{1}{2p^2k^2}\,(p^2+k^2)\,[\,p^2k^2+m^2\,k\cdot p\,]
\Bigg\}\nonumber\\ \nonumber\\
&+&\frac{\alpha}{8\pi\Delta^2}\left(\xi-1\right)\Bigg\{
\left[\frac{(p^2k^2-m^4)}{2}\,-\,\frac{(p^2-k^2)^2}{4}\,-\,\frac{3}{8\Delta^2}
(p^2-k^2)^2\,(p^2k^2-m^4)\right]{\it J}_{0}\nonumber\\
&& \hspace{25mm}
+\,\frac{m^2}{2p^2k^2}\,k\cdot p\,(\,k^2+\,p^2)\,-\,\frac{(p^2+k^2)}{2}\,-\,
\left(\,k\cdot p-\,m^2\right)\nonumber\\
&&\hspace{20mm}
+\,\frac{1}{\chi}\Bigg[
-\,m^2\left\{\left(p^2+\,m^2\,\frac{k^2}{p^2}\right){\it L}\,+\,
\left(\,k^2+\,m^2\,\frac{p^2}{k^2}\right){\it L^{'}}\right\}\,\Delta^2
\nonumber\\
&& \hspace{25mm}
-\,\frac{m^6}{2}q^2\,k\cdot p\,\left(\,\frac{\it L}{p^2}+\frac{\it L^{'}}{k^2}
\right)-\,\frac{1}{2}p^2k^2q^2\,k\cdot p\,\left(\,{\it L}+\,{\it L^{'}}\right)
\nonumber\\
&& \hspace{25mm}
+\frac{1}{4}q^2m^2\left[p^2(p^2-m^2){\it L}+k^2(k^2-
m^2){\it L^{'}}\right]
\nonumber\\
&& \hspace{25mm}
-\frac{1}{4}q^2m^2\left[p^2(k^2+m^2){\it L}+k^2(p^2+m^2){\it L^{'}}\right]
\nonumber\\
&& \hspace{25mm}
+\frac{1}{2}(p^2-k^2)q^2(m^4+p^2k^2)(\it L-\it L^{'})\nonumber\\
&& \hspace{25mm}
+\frac{m^2}{4}(p^4-k^4)\left[(p^2-m^2){\it L}-\,(k^2-m^2){\it L^{'}}\right]
\nonumber\\
&& \hspace{25mm}
+\frac{m^2}{2}(p^4-k^4)[k\cdot p-m^2]({\it L}-{\it L}^{'})
+m^2(p^2-k^2)(p^4{\it L}-k^4{\it L}^{'})\nonumber\\
&& \hspace{25mm}
-\frac{m^4}{4}(p^2-k^2)[p^2+2\,k\cdot p\,+k^2](\it L-\it L^{'})\nonumber \\
&& \hspace{25mm}
+\frac{3}{8\Delta^2}q^2(p^2-k^2)[p^2+2\,k\cdot p\,
+k^2]\left[(p^4k^2-m^6){\it L}-(k^4p^2-m^6){\it L^{'}}\right]\nonumber\\
&& \hspace{25mm}
-\frac{3m^2}{8\Delta^2}(p^2-k^2)^3\left[p^2(m^2-k^2){\it L}
-k^2(m^2-p^2){\it L^{'}}\right]\nonumber\\
&& \hspace{25mm}
+\frac{3m^4}{8\Delta^2}q^2(p^2-k^2)^2\left[(p^2-m^2){\it L}
+(k^2-m^2){\it L^{'}}\right]\nonumber\\
&& \hspace{25mm}
-\frac{3}{8\Delta^2}p^2k^2q^2(p^2-k^2)^2\left[(p^2-m^2){\it L}
+(k^2-m^2){\it L^{'}}\right]\nonumber\\
&& \hspace{22mm}
+\Bigg[
-8m^4\Delta^2+4m^2(p^2+k^2)\Delta^2+2m^2(p^2-k^2)^2[m^2-(p^2+k^2)]\nonumber\\
&& \hspace{25mm}
-2m^4q^2[m^2+\,k\cdot p\,]+2p^2k^2q^2[m^2+k\cdot
p\,]+2m^4(p^2-k^2)^2\nonumber\\
&& \hspace{25mm}
+\frac{3}{2\Delta^2}q^2(p^2-k^2)^2(m^4-p^2k^2)\left[m^2+k\cdot p\right]
\Bigg]{\it S}
\Bigg]
\Bigg\} \qquad ,
\end{eqnarray}
\newpage
\begin{eqnarray}
\tau_{4}=\frac{{\alpha}m}{8\pi\Delta^2}(\xi-1)\Bigg\{
&-&\left [1+\frac{3q^2}{2\Delta^2}\left[m^2+k\cdot p\,\right]\right ]{\it
J}_{0}
-\frac{2}{p^2k^2}\,k\cdot p\nonumber\\
+\frac{1}{\chi}\Bigg[
&-&\frac{4m^2}{(p^2-k^2)}\left(\frac{k^2}{p^2}{\it L}
-\frac{p^2}{k^2}{\it L^{'}}\right)
\Delta^2 +3m^2(p^2-k^2)({\it L}-{\it L^{'}})\nonumber\\
&-&\frac{m^4q^2}{(p^2-k^2)}[k^2+2\,k\cdot p\,+p^2]\left(\frac{{\it L}}{p^2}-
\frac{{\it L}^{'}}{k^2}\right)\nonumber\\
&-&q^2\left\{p^2\left(\frac{m^4}{p^4}-1\right){\it L}
+k^2\left(\frac{m^4}{k^4}-1\right){\it L^{'}}\right\}\nonumber\\
&+&\frac{3m^2q^2}{\Delta^2}(p^2-k^2)\left\{(m^2+p^2){\it L}
-(m^2+k^2){\it L^{'}}\right\}\nonumber\\
&-&\frac{3q^2}{2\Delta^2}(p^2-k^2)(m^4-p^2k^2)({\it L}-{\it L^{'}})
-\frac{3q^4}{2\Delta^2}(m^4-k^2p^2)({\it L}+{\it L^{'}})\nonumber\\
&-&\frac{3m^2q^4}{\Delta^2}\left\{(p^2-m^2){\it L}+(k^2-m^2){\it L^{'}}\right\}
\nonumber\\
&+&\Big\{-20m^2q^2-2q^2(p^2+k^2)-\frac{12m^2q^4}{\Delta^2}[m^2+k\cdot p\,]
\nonumber\\
&+&\frac{6q^4}{\Delta^2}(m^4-p^2k^2)\Big\}{\it S}
\Bigg]
\Bigg\}\qquad\qquad\qquad ,
\end{eqnarray}
\vspace{3mm}
\begin{eqnarray}
\tau_{5}=\frac{{\alpha}m}{8\pi\Delta^2}(\xi-1)\Bigg\{
&-&\left[\Delta^2-\frac{1}{4}(p^2-k^2)^2+\frac{q^2}{2}[m^2+k\cdot p\,]
\right]{\it J}_0+\frac{(p^2+k^2)}{p^2k^2}\Delta^2\nonumber\\
+\frac{1}{\chi}\Bigg[
&-&m^2(p^2-k^2)\left(\frac{p^2}{k^2}{\it L^{'}}-\frac{k^2}{p^2}{\it
L}\right)\Delta^2+2m^2(p^2-k^2)({\it L}-{\it L^{'}})\Delta^2\nonumber\\
&+&2m^2q^2\left\{\left(1-\frac{m^2}{p^2}\right){\it L}
+\left(1-\frac{m^2}{k^2}\right){\it L^{'}}\right\}\Delta^2\nonumber\\
&+&m^2q^2\left\{(m^2+k^2)\frac{{\it L}}{p^2}
+(m^2+p^2)\frac{{\it L}^{'}}{k^2}\right\}\Delta^2\nonumber\\
&+&\frac{m^2q^2}{2}\left\{q^2[m^2+\,k\cdot p\,]+q^2\,k\cdot p\,
-\frac{1}{2}(p^2-k^2)^2\right\}({\it L}+{\it L^{'}})\nonumber\\
&+&\frac{m^2}{2}(p^2-k^2)\Bigg\{q^2m^2+2q^2\,k\cdot p\,
-\frac{1}{2}(p^2-k^2)^2\Bigg\}({\it L}-{\it L^{'}})\nonumber\\
&+&\frac{1}{4}q^4(p^2+k^2)(p^2{\it L}+k^2{\it L^{'}})
-\frac{1}{4}q^2(p^4-k^4)(p^2{\it L}-k^2{\it L^{'}})\nonumber\\
&+&q^2(p^2+k^2-2m^2)\left(q^2 m^2+q^2 k\cdot p
+2\Delta^2\right){\it S}
\Bigg]
\Bigg\}\qquad ,
\end{eqnarray}
\newpage
\begin{eqnarray}
\tau_{6}&=&\frac{(p^2-k^2)}{2}\tau_{2}(\xi=1) \hspace{80mm} \nonumber\\
\nonumber\\
&+&\frac{\alpha}{8\pi\Delta^2}(\xi-1)\Bigg\{
(p^2-k^2)\left[\frac{q^2}{4}
-\frac{3q^2}{8\Delta^2}(m^4-p^2k^2)\right]{\it J}_0
-\frac{(p^2-k^2)}{2p^2k^2}\left[m^2\,k\cdot p\,-p^2k^2\right]\nonumber\\
\nonumber \\
&& \hspace{15mm}
+\frac{1}{\chi}\Bigg[
m^2\Delta^2\left\{\left(p^2-m^2\frac{k^2}{p^2}\right){\it L}
-\left(k^2-m^2\frac{p^2}{k^2}\right){\it L^{'}}\right\}\nonumber\\\nonumber \\
&& \hspace{20mm}
+\frac{1}{2}m^2(k^4-p^4)[m^2-k\cdot p\,]({\it L}+{\it L^{'}})
-2m^2(k\cdot p)(p^4{\it L}-k^4{\it L^{'}})\nonumber\\ \nonumber \\
&& \hspace{20mm}
-\frac{m^6q^2}{2}\left\{\left(1+\frac{k\cdot p\,}{p^2}\right){\it L}
-\left(1+\frac{k\cdot p}{k^2}\right){\it L^{'}}\right\}\nonumber\\\nonumber \\
&& \hspace{20mm}
-\frac{m^2}{2}\,k\cdot p\,(p^2-k^2)\left[(m^2-p^2){\it L}+(m^2-k^2){\it L^{'}}
\right]\nonumber\\
&& \hspace{20mm}
-\frac{1}{2}p^2k^2q^2\left\{\left[k^2-k\cdot p\,\right]{\it L}
-\left[p^2-k\cdot p\,\right]{\it L^{'}}\right\}\nonumber\\
&& \hspace{20mm}
-q^2\left[p^4(m^2-k^2){\it L}-k^4(m^2-p^2){\it L^{'}}\right]
+2m^2k^2p^2\left(k^2{\it L}-p^2{\it L^{'}}\right)\nonumber\\
&& \hspace{20mm}
+\frac{3m^2}{8\Delta^2}(p^4-k^4)(p^2-k^2)
\left[p^2(m^2-k^2){\it L}-k^2(m^2-p^2){\it L^{'}}\right]\nonumber\\
&& \hspace{20mm}
-\frac{3m^2}{4\Delta^2}p^2k^2q^2(p^2-k^2)
\left[(m^2+p^2){\it L}+(m^2+k^2){\it L^{'}}\right]\nonumber\\
&& \hspace{20mm}
-\frac{3m^2}{4\Delta^2}p^2k^2(p^2-k^2)^2
\left[(m^2-p^2){\it L}-(m^2-k^2){\it L^{'}}\right]\nonumber\\
&& \hspace{20mm}
-\frac{3m^2}{8\Delta^2}p^2q^2(p^4-k^4)
\left[(m^2+k^2){\it L}-(m^2+p^2){\it L^{'}}\right]\nonumber\\
&& \hspace{20mm}
+\frac{3}{8\Delta^2}q^4(p^2-k^2)
\left[(m^6+p^4k^2){\it L}+(m^6+p^2k^4){\it L^{'}}\right]\nonumber\\
&& \hspace{20mm}
+\frac{3q^2}{8\Delta^2}(p^2-k^2)^2
\left[(m^6-p^4k^2){\it L}-(m^6-p^2k^4){\it L^{'}}\right]\nonumber\\\nonumber\\
&& \hspace{20mm}
+(p^2-k^2)\Bigg[
-\frac{3m^4}{2\Delta^2}q^4\left[m^2+\,k\cdot p\,\right]
+\frac{3}{2\Delta^2}k^2p^2q^4\left[m^2+\,k\cdot p\,\right]\nonumber\\
&& \hspace{40mm}
-4m^4q^2+2m^2q^2(p^2+k^2)
\Bigg]\,\,{\it S}\,\,
\Bigg]
\Bigg\}\qquad\qquad ,
\end{eqnarray}
\newpage
\begin{eqnarray}
\tau_{7}=\frac{\alpha m}{8\pi\Delta^2}(\xi-1)\Bigg\{
&-&\frac{q^2}{2}{\it J}_{0}-\frac{2}{p^2k^2}\Delta^2 \hspace{80mm}\nonumber\\
\nonumber\\
\
&+&\frac{1}{\chi}\Bigg[
\frac{-2m^2q^2}{(p^2-k^2)}\left\{\frac{(m^2-k^2)}{p^2}{\it L}
-\frac{(m^2-p^2)}{k^2}{\it L^{'}}\right\}\Delta^2 \nonumber\\\nonumber \\
&& \hspace{5mm}
-2m^2\left\{\left(1-\frac{k^2}{p^2}\right){\it L}+
\left(1-\frac{p^2}{k^2}\right){\it L^{'}}\right\}\Delta^2\nonumber\\\nonumber
\\
&& \hspace{5mm}
-q^2(k\cdot p)\left[(m^2-p^2){\it L}+(m^2-k^2){\it L^{'}}\right]\nonumber\\
&& \hspace{5mm}
+q^2\left[p^2(m^2-k^2){\it L}+k^2(m^2-p^2){\it L^{'}}\right]\nonumber\\
&& \hspace{5mm}
-2q^2\left\{q^2\left[m^2+k\cdot p\right]+2\Delta^2\right\}\,\,{\it S}\,\,
\Bigg]
\Bigg\} \qquad ,
\end{eqnarray}
\noindent and
\vspace{5mm}
\begin{eqnarray}
\tau_{8}\,=\,\frac{\alpha}{4\pi\Delta^2}\Bigg\{
\frac{q^2}{2}\left[k\cdot p\,+m^2\right]{\it J}_{0}
+2q^2{\it S}
+\left[-p^2+k\cdot p\,\right]{\it L}
+\left[-k^2+k\cdot p\,\right]{\it L^{'}}
\Bigg\}\qquad.
\end{eqnarray}
These ${\tau_i}$'s are given in an arbitrary covariant gauge specified by
${\xi}$. The Ball-Chiu calculation was performed in the Feynman gauge
${(\xi=1)}$. If we compare our ${\tau}$'s with the result of Ball-Chiu,
then their result needs a few corrections:
\begin{itemize}
\item
in Eq.~(3.8e) there should be $(+)$ sign instead $(-)$ front of the second
term.

\item
in Eq.~(3.8e) ${\frac{p\cdot p^{'}}{2}\left(\frac{1-m^2}{p^2}\right){\it L}}$
 should be  ${\frac{p\cdot p^{'}}{2}\left(1-\frac{m^2}{p^2}\right){\it L}}$.

\item
in Eq.~(3.11b) the third term with ${\ln(p^2/p^{'2})}$ should have 1/2 factor
in front.

\item
in Eq.~(3.11c) the third coefficient ${I_0}$, ${(p^2-p^{'2})^2/8}$ should be
${(p^2+p^{'2})^2/8}$.

\item
in Eq.~(3.12) there is an overall factor 2 missing.

\item
in Eq.~(3.14) factor 1/6 should be 1/12.

\item
in Eq.~(A18) the coefficient of the first term should not have a factor of 1/2.

\item
in Eq.~(A.19) first coefficient ${p\cdot p^{'}/2}$ should be
${p\cdot p^{'}}$.

\end{itemize}
where the equation numbers refer to those in~\cite{BC}.
\section{Freedom from Kinematic Singularities}
Clearly the full vertex, ${\Gamma^{\mu}(k,p)}$, is free of kinematic
singularities. The Ball-Chiu construction ensures that the longitudinal part
is free, so the transverse part must be. However, after decomposing this
transverse part into 8 components, it is not necessary that the
individual components will each be free of kinematic singularities. Ball
and Chiu showed that with their choice of eight basis vectors, the
transverse vertex in the Feynman gauge possessed this property of being
singularity free. Here we explicitly consider whether this is true in
an arbitrary
covariant gauge. Indeed such checks are far longer than the initial
calculation reported above. We consider several limits in turn.
{\noindent}
\subsection{${\Delta^2\rightarrow 0}:$}

    The proof depends crucially on the behaviour of
the combination of Spence functions forming the  integral ${{\it J}_0}$
that appears in every ${\tau_{i}}$. Thus, for instance, when we consider the
limit ${\Delta^2\rightarrow 0}$, i.e. ${(k\cdot p)^2\rightarrow k^2p^2}$,
we can  deduce
from Eqs.~(26,A15-A19) that ${{\it J}_0}$ can be expanded in powers of
${\Delta^2}$ as:
\begin{eqnarray}
{\it J}_{0}={\it J}_0^0+{\it J}_0^1\Delta^2 +{\cal O}(\Delta^4)
\end{eqnarray}
where
\def\pk{k\cdot p}
\def\wort{\sqrt{k^2p^2}}
\def\cf{\frac{i\pi^2}{2}}
\def\[{\left[}
\def\]{\right]}
\begin{eqnarray}
{\it J}_0^0&=&
- \frac{1}{m^2+\wort}\Bigg[4 S
-2\frac {(k^2+\wort)}{k^2-p^2} L'
+2\frac{(p^2+\wort)}{k^2-p^2}L\Bigg]\\
{\it J}_0^1&=&\Bigg[
\frac{2}{3q_0^2(m^2+\wort)\wort}
-Y_1(k^2,p^2) L'-Y_1(p^2,k^2) L
-Z_1(k^2,p^2) S
\Bigg]
\end{eqnarray}
and $Y_1$ and $Z_1$ are defined as
\begin{eqnarray}
Y_1(k^2,p^2)&=&\frac{\left(k^2-m^2\right)}{3\left(\sqrt{k^2}
-\sqrt{p^2}\right)^3\left(m^2+\wort\right)^3\wort}\Bigg[
3\sqrt{k^2}\left(m^2-p^2\right)\nonumber\\
&+&\sqrt{p^2}\left(k^2-m^2\right)
\Bigg]\\\nonumber\\
Z_1(k^2,p^2)&=&-
\frac{1}{q_0^2\left(m^2+\wort\right)^3\left(q_0^2-4m^2\right)\wort}\Bigg[
8m^6-8m^4\left(k^2+p^2-\frac{4}{3}\wort\right)\nonumber\\
&+&m^2\left(2 q_0^4+\frac{8}{3}\wort\left(k^2+p^2+\wort\right)\right)
+\frac{2}{3}\wort q_0^4
\Bigg],\\\nonumber\\
q_0^2&=&k^2+p^2-2\wort.
\end{eqnarray}
Together with the known behaviour of all the other functions,
such as ${\it L}$,
${L^{'}}$ and ${\it S}$, it is a lengthy but straightforward calculation to
deduce
 that each ${\tau_{i}}$ is finite in  the limit ${\Delta^2 \rightarrow 0}$,
despite the appearance of explicit ${1/\Delta^2}$ and ${1/\Delta^4}$ terms. \\

\subsection{\bf{$\chi \rightarrow 0$:}}

    As seen from Eq.~(20) the full vertex, and hence the transverse part,
has no pole singularities when ${\chi \rightarrow 0}$. However, the expressions
for ${\tau_2, \tau_3, \tau_4, \tau_5, \tau_6}$ and ${\tau_7}$, Eqs.~(68-73),
have explicit factors of ${1/\chi}$ in all but the Feynman gauge. As can be
seen from Eq.~(55) ${\chi}$
only vanishes if ${\bf both}$ ${p^2}$ and $k^2$ tend to $ m^2$, i.e.
when both of the fermion legs, Fig.~1, are on-mass-shell, then when
${k^2 \rightarrow p^2}$, ${\chi = (q^2-4m^2)(p^2-m^2)^2}$.
In this limit
the full vertex only has logarithmic singularities, like
${\ln\left(1-\frac{m^2}{p^2}\right)}$. Consequently, an acceptable basis
for the
transverse vertex is one in which only these logarithmic singularities occur.
Explicit calculation shows that ${\tau_2, \tau_3, \tau_5}$ and ${\tau_6}$,
given by Eqs.~(68,69,71,72), do have only these logarithmic terms when
${\chi \rightarrow 0}$. However, both ${\tau_4}$ and ${\tau_7}$ have poles
in ${1/(p^2-m^2)}$ term.
These singularities
are readily removed by choosing a new basis for the transverse vertex, the
${S_i^{\mu}}$ ${(i=1,....,8)}$.
Clearly this only involves changes to ${T^{\mu}_4}$ and ${T^{\mu}_7}$. Note
that these singularities do not arise in the Feynman gauge ${(\xi=1)}$, and
so Ball and Chiu were not aware of this constraint.\\

   We write
\begin{eqnarray}
\Gamma^{\mu}_T(k,p)=\sum^8_{i=1}\,\sigma^iS^{\mu}_i
\end{eqnarray}
where
\begin{eqnarray}
S^{\mu}_i&\equiv& T^{\mu}_i\;\;\;\;\;   \mbox{for} \;\;\;\;\;i=1,2,3,5,6,7,8
\end{eqnarray}
%
%
and
\begin{eqnarray}
\hspace{23mm}
S^{\mu}_4&=&q^2\left[\gamma^{\mu}(\not\!p+\not\!k)-p^{\mu}-k^{\mu}\right]
+2(p-k)^{\mu}k^{\lambda}p^{\nu}\sigma_{\lambda\nu}
\end{eqnarray}
then
\begin{eqnarray}
\sigma_i&\equiv&\tau_i\;\;\;\;\; \mbox{for} \;\;\;\;\;i=1,2,3,5,6,8
\end{eqnarray}
and
\begin{eqnarray}
\sigma_4&=&\frac{(k^2-p^2)}{4}\tau_4\\
\sigma_7&=&\tau_7+\frac{q^2}{2}\tau_4 \qquad .
\end{eqnarray}
${\sigma_7}$ is then given explicitly as:
\begin{eqnarray}
\sigma_7&=&\frac{{\alpha}m}{8\pi\Delta^2}(\xi-1)\Bigg\{
\left(-q^2-\frac{3q^4}{4\Delta^2}(m^2+k\cdot p)\right){\it J}_0
-\frac{q^2}{p^2k^2}\,k\cdot p-2\frac{\Delta^2}{p^2k^2}\nonumber\\
&&\hspace{5mm}
+\frac{1}{\chi}\Bigg[
\Delta^2\left\{
-2\frac{m^4q^2}{(p^2-k^2)}\left(\frac{{\it L}}{p^2}-
\frac{{\it L}^{'}}{k^2}\right)-2m^2(p^2-k^2)\left(\frac{{\it L}}{p^2}-
\frac{{\it L}^{'}}{k^2}\right)
\right\}\nonumber\\
&&\hspace{10mm}
-\frac{m^4q^4}{2(p^2-k^2)}[k^2+2(k\cdot p)+p^2]\left(\frac{{\it L}}{p^2}-
\frac{{\it L}^{'}}{k^2}\right)\nonumber\\
&&\hspace{10mm}
-\frac{q^4}{2}\left\{p^2\left(\frac{m^4}{p^4}-1\right){\it L}
+k^2\left(\frac{m^4}{k^4}-1\right){\it L^{'}}\right\}\nonumber\\
&&\hspace{10mm}
+\frac{3}{2}\frac{m^2q^4}{\Delta^2}(p^2-k^2)\left\{(m^2+p^2){\it L}
-(m^2+k^2){\it L^{'}}\right\}\nonumber\\
&&\hspace{10mm}
-\frac{3q^4}{4\Delta^2}(p^2-k^2)(m^4-p^2k^2)({\it L}-{\it L}^{'})\nonumber\\
&&\hspace{10mm}
-\frac{3q^6}{4\Delta^2}(m^4-p^2k^2)({\it L}+{\it L}^{'})
+\frac{3m^2q^2}{2}(p^2-k^2)\left({\it L}-{\it L}^{'}\right)\nonumber\\
&&\hspace{10mm}
-\frac{3m^2q^6}{2\Delta^2}\left[(p^2-m^2){\it L}+(k^2-m^2){\it L}^{'}\right]
\nonumber\\
&&\hspace{10mm}
+q^2k\cdot p\left[(p^2-m^2){\it L}+(k^2-m^2){\it L}^{'}\right]\nonumber\\
&&\hspace{10mm}
-q^2\left[p^2(k^2-m^2){\it L}+k^2(p^2-m^2){\it L}^{'}\right]\nonumber\\
&&\hspace{10mm}
+\Big\{
-10m^2q^4-q^4(p^2+k^2)-6\frac{m^2q^6}{\Delta^2}(m^2+k\cdot p)
+\frac{3q^6}{\Delta^2}(m^4-p^2k^2)\nonumber\\
&&\hspace{10mm}
-2q^4(m^2+k\cdot p)-4q^2\Delta^2\Big\}{\it S}
\Bigg]
\Bigg\}
\end{eqnarray}
In this new basis, all the ${\sigma_i}$'s ${(i=1,....,8)}$ have no
singularities
other than the expected logarithmic ones. Note that in this new basis, the
C-parity operation of Eq.~(15) requires
\begin{eqnarray}
\sigma_4(k^2,p^2,q^2)=-\sigma_4(p^2,k^2,q^2)
\end{eqnarray}
which Eq.~(83) ensures.
\subsection{Asymptotic limit:}

   It is convenient to give here the simple asymptotic limit for the
transverse vertex. In the limit that either of the fermion momenta are
large, e.g. ${k^2 >> p^2 >> m^2}$,
\begin{eqnarray}
{\it J}_{0}=\frac{2}{k^2}\left[\left(1+\frac{(k\cdot p)}{k^2}\right)
\ln \frac{k^2}{p^2}
+2\left(1+\frac{(k\cdot p)}{k^2}\right)\right]+O(1/k^4)
\end{eqnarray}
Consequently, the transverse vertex has the following well known limit
\begin{eqnarray}
\Gamma^{\mu}_{T}=\frac{\alpha\xi}{8\,\pi}
\left[\frac{k^{\mu}{\not \! k}}{k^2}
-\gamma^{\mu}\right] \ln \left(\frac{k^2}{p^2}\right)
\end{eqnarray}
\section{Conclusions}
\indent
    This paper presents the complete one loop calculation of the
fermion-boson vertex in QED in an arbitrary covariant gauge. This
calculation has, in fact, been performed independently in Durham and
Groningen. The authors have joined forces only to compare and  check
their answer and to write this paper. The coupling
of two spin-1/2 particles with a vector boson involves twelve independent
spin and Lorentz vectors. Each of these vectors has a coefficient that is
an analytic function of the three Lorentz scalars, $k^2$, $p^2$ and $q^2$,
that can be formed from the two independent 4-momenta flowing through the
vertex. These twelve components are given as functions of the covariant
gauge parameter. They have been previously calculated by Ball and
Chiu~\cite{BC}
in the Feynman gauge. Our results correct some typographical errors in
their publication. The vertex has only logarithmic singularities: these
arise either when the external legs are on-shell or when the internal
fermions can be real.

    Four of the 12 components define what is called the longitudinal vertex.
This is related by the Ward-Takahashi identity to the fermion propagator.
This fact allows three of these components to be expressed in terms of the
fermion wavefunction renormalization $F(p^2)$, and its mass function
$M(p^2)$ and forces a fourth to be zero. Ball and Chiu have shown how to
construct this longitudinal vertex in
a way free of kinematic singularities. This freedom is essential in
ensuring the Ward identity is the ${q\rightarrow 0}$ limit of the
Ward-Takahashi identity. Subtraction of this longitudinal vertex from our
one loop answer leaves the transverse vertex to ${\cal O(\alpha)}$. This can
be represented in terms of a basis of eight vectors orthogonal to the boson
momentum, each unconstrained by the Ward-Takahashi identity.

    We propose a new transverse basis ${S_i^{\mu}}$ ${\left(i=1...,8\right)}$,
Eqs.~(82,83), which has components with only the logarithmic singularities
of the full vertex. This basis modifies the ${T_i^{\mu}\, (i=1,....,8)}$
of Eq.~(13) proposed by Ball and Chiu~\cite{BC}. Though their basis has no
additional singularities in the Feynman gauge, this is not the case in
any other gauge. Eqs.~(19,66-74,81-87) constitute our new result in QED to
one loop.
The same and/or related integrals arise in QCD, and so this
calculation  could, in principle, be extended to non-Abelian theories in
any covariant gauge too.

    Though our calculations are self-evidently only true to ${\cal O(\alpha)}$,
our aim is wider. The hope is that the coefficients of each of the transverse
vectors, ${S_i^{\mu}}$, like those of the longitudinal component, are free of
kinematic singularities at all orders in perturbation theory and even
non-perturbatively. Just as use of the Ward-Takahashi identity specifies
non-perturbatively the longitudinal vertex in terms of the fermion
propagator, Eq.~(9),
multiplicative renormalizability too imposes relationships between the
vertex and the fermion propagator. These constrain the transverse vertex.
A start has been made in analysing these powerful conditions. Ignoring such
requirements and use of, for instance, a bare vertex ({\it{the rainbow
approximation}}) in studies of chiral symmetry breaking leads to the generation
of highly gauge dependent masses. In contrast non-perturbative enforcement of
the Ward-Takahashi identity and the constraints of multiplicative
renormalizability dramatically reduces or even eliminates~\cite{CP2,BP} this
unphysical gauge dependence. Indeed, knowing the vertex in any covariant gauge
may give us an understanding of how the essential gauge dependence of the
vertex demanded by its Landau-Khalatnikov transformation~\cite{LK} is
satisfied non-perturbatively. Moreover, having a basis for the transverse
vertex with
coefficients free of non-dynamical singularities is a key step in further
investigations of a meaningful non-perturbative truncation of
Schwinger-Dyson equations. For instance, studying the behaviour of just the
propagators one must have an Ansatz for the 3-point vertex that embodies
as completely as possible the constraints of gauge invariance on all the
higher $n$-point functions. Fulfilling the Ward-Takahashi identity and
multiplicative renormalizability are essential in constructing such an
Ansatz. Moreover in the weak coupling limit this vertex must agree with
perturbation theory --- hence this one loop calculation.
\vskip 2cm

\noindent{\bf{Acknowledgements}}
{}~~A.K. and M.R.P. are grateful to Adnan Bashir
 for many helpful comments.  M.R. wishes
to thank David Atkinson, Valery Gusynin and Porter Johnson for useful remarks
and
 interesting discussions.  A.K. would also like to thank the University
 of Istanbul for
 a scholarship that made this work possible.
\newpage
\setcounter{section}{0}
\renewcommand{\thesection}{Appendix \Alph{section}}
\section{}
\renewcommand{\thesection}{\Alph{section}}
\setcounter{equation}{0}
\renewcommand{\theequation}{\thesection.\arabic{equation}}

\baselineskip=6.8mm
\vskip 5mm
\noindent Here we collect the results of evaluating all the integrals necessary
for our calculation of Sect.~2.2 in ${d=4+\epsilon}$ dimensions.
\begin{eqnarray}
Q_{1}&=&\int_{M}d^{d}k\, \frac{1}{(k-p)^2\,\left[k^2-m^2\right]}
\nonumber\\
&=&{\it
i}\pi^2{\mu^{\epsilon}}\left\{C-\left(1-\frac{m^2}{p^2}\right)\ln\left(1-\frac{p^2}{m^2}\right)\right\}\qquad,\\\nonumber\\
Q_{2}^{\nu}&=&\int_{M}d^{d}k\,\frac{k^{\nu}}{(k-p)^2\,
\left[k^2-m^2\right]}\nonumber\\
&=&{\it i}\pi^2{\mu^{\epsilon}}\left\{\frac{p^{\nu}}{2}\left
[C+2+\frac{m^2}{p^2}-\left(1-\frac{m^4}{p^4}
\right)\ln\left(1-\frac{p^2}{m^2}\right)\right]\right\} \qquad,\\\nonumber\\
Q_{3}&=&\int_{M}d^{d}k\, \frac{1}{(k-p)^4\,\left[k^2-m^2\right]}\nonumber\\
&=&{\mu^{\epsilon}}\frac{{\it i}\pi^2}{(m^2-p^2)}\left\{
C-\left(1+\frac{m^2}{p^2}\right)
\ln\left(1-\frac{p^2}{m^2}\right)\right\} \qquad,\\\nonumber\\
Q_{4}^{\nu}&=&\int_{M}d^{d}k\,\frac{k^{\nu}}{(k-p)^4\,
\left[k^2-m^2\right]}\nonumber\\
&=&{\mu^{\epsilon}}\frac{{\it i}\pi^2p^{\nu}}{(m^2-p^2)}\left\{
C+\left(1-\frac{m^2}{p^2}\right)-\left(1+\frac{m^4}{p^4}\right)
\ln\left(1-\frac{p^2}{m^2}\right)\right\} \qquad,\\\nonumber\\
Q_{5}&=&\int_{M}d^{d}k\,\frac{k^2}{(k-p)^4\,\left[k^2-m^2\right]}\nonumber\\
&=&m^2Q_{3} \qquad,\\\nonumber\\
Q_{6}^{\nu}&=&\int_{M}d^{d}k\,\frac{k^2\,k^{\nu}}{(k-p)^4\,
\left[k^2-m^2\right]}\nonumber\\
&=&m^2Q_{4} \qquad,\\\nonumber\\
Q_{7}&=&\int_{M}d^{d}w\,\frac{1}{(k-w)^2-m^2\,
\left[(p-w)^2-m^2\right]}\nonumber\\
&=&{\it i}\pi^2{\mu^{\epsilon}}\left[C+2-2{\it S}\right] \qquad,\\\nonumber\\
Q_{8}&=&\int_{M}d^{d}w\,\frac{1}{\left[(p-w)^2-m^2\right]\,w^2}\nonumber\\
&=&{\it i} \pi^2{\mu^{\epsilon}}\left[C+2-{\it L}\right] \qquad,\\\nonumber\\
Q_{9}^{\nu}&=&\int_{M}d^{d}w\,\frac{w^{\nu}}{\left[(k-w)^2-m^2\right]\,
\left[(p-w)^2-m^2\right]}\nonumber\\
&=&\frac{{\it i}\pi^2}{2}{\mu^{\epsilon}}(p^{\nu}+k^{\nu})\left[
C+2-2{\it S}\right] \qquad,\\ \nonumber\\
Q_{10}^{\nu}&=&\int_{M}d^{d}w\,\frac{w^{\nu}}{\left[(p-w)^2-m^2\right]\,w^2}
\nonumber\\
&=&\frac{{\it i}\pi^2}{2}{\mu^{\epsilon}}p^{\nu}\left\{C+2-\frac{m^2}{p^2}
-\left(1-\frac{m^2}{p^2}\right)\ln\left(1-\frac{p^2}{m^2}\right)\right\}
\qquad,\\\nonumber\\
Q_{11}&=&\int_{M}d^{d}w\,\frac{1}{\left[(p-w)^2-m^2\right]\,w^4}\nonumber\\
&=&\frac{{\it i}\pi^2}{(m^2-p^2)}{\mu^{\epsilon}}\left\{C-\left(
1+\frac{m^2}{p^2}\right)
\ln\left(1-\frac{p^2}{m^2}\right)\right\} \qquad,\\\nonumber\\
Q_{12}^{\nu}&=&\int_{M}d^{d}w\,\frac{w^{\nu}}{\left[(p-w)^2-m^2\right]\,w^4}
\nonumber\\
&=&{\it i}\pi^2\frac{p^{\nu}}{p^2}\left[1+\frac{m^2}{p^2}
\ln\left(1-\frac{p^2}{m^2}\right)\right]\qquad,\\\nonumber\\
Q_{13}&=&\int_{M}d^{d}w\,\frac{1}{\left[(p-w)^2-m^2\right]\,
\left[(k-w)^2-m^2\right]\,w^4}=I^{(0)}\nonumber \\\nonumber\\
&=&{\it i}\pi^2{\mu^{\epsilon}}\Bigg\{\frac{1}{\chi}\Bigg[-2q^2{\it S}
+p^2\frac{(p^2-m^2)q^2+2m^2(k^2-p^2)}{(p^2-m^2)^2}{\it L} \qquad\nonumber\\
\nonumber\\
&& \hspace{15mm}
+k^2\frac{(k^2-m^2)q^2-2m^2(k^2-p^2)}{(k^2-m^2)^2}{\it L^{'}}\Bigg]
-\frac{C}{(p^2-m^2)(k^2-m^2)}
\Bigg\} \qquad\nonumber\\
\end{eqnarray}
\noindent where we recall\\
\begin{eqnarray}
C&=&-\frac{2}{\epsilon}-\gamma-\ln(\pi)-\ln\frac{m^2}{\mu^2},\qquad\nonumber\\
\nonumber\\
{\it L}&=&\left(1-\frac{m^2}{p^2}\right)\,\ln \left(1-\frac{p^2}{m^2}\right)
\qquad ,\nonumber\\
{\it L}^{'}&=&{\it L}\left(p\leftrightarrow k\right)\qquad ,\nonumber\\
{\it S}&=&\frac{1}{2}\left(1-4\,\frac{m^2}{q^2}\right)^{1/2}\,
\ln \frac{\left[\left(1-{4m^2}/{q^2}\right)^{1/2}+1\right]}
{\left[\left(1-{4m^2}/{q^2}\right)^{1/2}-1\right]}\qquad .\qquad\;
\end{eqnarray}
\begin{eqnarray*}
Q_{14}&=&\int_{M}d^{d}w\,\frac{1}{\left[(p-w)^2-m^2\right]\,
\left[(k-w)^2-m^2\right]\,w^2}=J^{(0)}\nonumber\\\nonumber \\
\end{eqnarray*}
$J^{(0)}$ is naturally expressed in terms of the Spence function $Sp(x)$~:
\begin{eqnarray}
Sp(x)&=&-\int_0^x\,dy\frac{\ln(1-y)}{y}\qquad,\quad\qquad
\end{eqnarray}
\noindent so that\\
\begin{eqnarray}
J^{(0)}&=&\frac{i\pi^2}{-2\,\Delta}\Bigg\{
Sp\left(\frac{y_{1}}{y_{1}-1}\right)
+Sp\left(\frac{y_1}{y_1-\frac{m^2}{p^2}}\right)
-Sp\left(\frac{y_1-1}{y_1-\frac{m^2}{p^2}}\right)\nonumber\\ \nonumber\\
&-&Sp\left(\frac{y_2}{y_2-1}\right)
-Sp\left(\frac{y_2}{y_2-\frac{m^2}{k^2}}\right)
+Sp\left(\frac{y_2-1}{y_2-\frac{m^2}{k^2}}\right)\nonumber\\\nonumber\\
&+&Sp\left(\frac{y_3}{y_3-q_1}\right)
-Sp\left(\frac{y_3-1}{y_{3}-q_1}\right)
+Sp\left(\frac{y_3}{y_3-q_2}\right)
-Sp\left(\frac{y_3-1}{y_{3}-q_2}\right)
\Bigg\}, \;\;\quad
\end{eqnarray}
\noindent where\\
\begin{eqnarray}
\alpha&=&1+\frac{-(k\cdot p)+\Delta}{p^2},\;\;
y_{1}=y_{0}+\alpha,\;\;
y_{2}=\frac{y_{0}}{(1-\alpha)},\;\;
y_{3}=-\frac{y_{0}}{\alpha} \qquad,\nonumber\\\nonumber\\
y_{0}&=&\frac{1}{2\,p^2\,\Delta}\left[k^2p^2-2(k\cdot p)^2
+2(k\cdot p)\Delta-p^2\Delta+p^2(k\cdot p)-m^2(k\cdot p-\Delta)\right]\qquad,
\nonumber\\\nonumber\\
q_1&=&\frac{1+\sqrt{1-4m^2/q^2}}{2},\;\;\;
q_2=\frac{1-\sqrt{1-4m^2/q^2}}{2} \qquad.\nonumber\\
\end{eqnarray}
In massless case, $J_0$ simplifies to\\
\begin{eqnarray}
J_0=\frac{2}{\Delta}\left[Sp\left(\frac{p^2-k\cdot p+\Delta}{p^2}\right)
-Sp\left(\frac{p^2-k\cdot p-\Delta}{p^2}\right)
+\frac{1}{2}\ln\left(\frac{k\cdot p-\Delta}{k\cdot p+\Delta}\right)
\ln\left(\frac{q^2}{p^2}\right)\right]
\end{eqnarray}
\newpage

\renewcommand{\thesection}{Appendix \Alph{section}}
\section{}
\renewcommand{\thesection}{\Alph{section}}
\setcounter{equation}{0}
\renewcommand{\theequation}{\thesection.\arabic{equation}}

\setcounter{equation}{0}
\baselineskip=7mm
\vskip 5mm
\noindent In this Appendix the coefficients of the 12 vectors ${V_i^{\mu}}$ in
Eq.~(6) are explicitly tabulated.
\begin{eqnarray}
a^{(1)}_{1}&=&3p^2(k^2-m^2)-2\,k\cdot p\,(p^2-m^2)+4\Delta^2,\nonumber\\
a^{(1)}_{2}&=&k\cdot
p\,(p^2+m^2)-\frac{3m^2p^2}{2}-\frac{p^2k^2}{2},\nonumber\\
a^{(2)}_{1}&=&k^2(k^2-m^2),\nonumber\\
a^{(2)}_{2}&=&k^2\,k\cdot p-\frac{k^2}{2}(m^2+k^2),\nonumber\\
a^{(3)}_{1}&=&k^2(p^2-m^2)-2\,k\cdot p\,(k^2-m^2)+4\Delta^2,\nonumber\\
a^{(3)}_{2}&=&m^2\,k\cdot p+\Delta^2-\frac{k^2}{2}(m^2+p^2),\nonumber\\
a^{(4)}_{1}&=&k^2(p^2-m^2)-2\,k\cdot p\,(k^2-m^2),\nonumber\\
a^{(4)}_{2}&=&
-\frac{3k^2p^2}{2}-\frac{m^2k^2}{2}+k\cdot p\,(m^2+k^2)-2\,\Delta^2,\nonumber\\
a^{(5)}_{1}&=&-2\Delta^2,\nonumber\\
a^{(5)}_{2}&=&-\frac{\Delta^2}{2},\nonumber\\
a^{(6)}_{1}&=&-(k^2+m^2)\,\Delta^2,\nonumber\\
a^{(6)}_{2}&=&-\left(\frac{m^2}{2}+k^2\right)\Delta^2,\nonumber\\
a^{(7)}_{1}&=&-8m\Delta^2,\nonumber\\
a^{(7)}_{2}&=&m\,\left[-3\,(k\cdot p)^2+k^2\,k\cdot p+2p^2\,k\cdot p
-2\Delta^2\right],\nonumber\\
%
%
a^{(8)}_{2}&=&
m\left[2k^2\,k\cdot p-\frac{k^2p^2}{2}-(k\cdot p)^2-\frac{k^4}{2}\right],
\nonumber\\
%
%
%
a^{(9)}_{2}&=&m\left[-k^2+k\cdot p\right],\nonumber\\
%
%
%
a^{(10)}_{2}&=&
m\left[-\frac{3p^2}{2}-\frac{k^2}{2}+2\,k\cdot p\right],\nonumber\\
%
%
%
a^{(11)}_{2}&=&-m\frac{q^2}{2}\,k\cdot p
\nonumber\\
%
%
%
a^{(12)}_{2}&=&m\frac{q^2}{2}\,k^2,\nonumber\\
a^{(i)}_1&=&0, \;\;\;\;\;\;\;\;\;\;i=8,9,10,11,12\qquad.
\end{eqnarray}
\newpage
\begin{eqnarray}
b_1^{(1)}&=&a_1^{(2)}(k \leftrightarrow p),\hspace{30mm}
b_1^{(2)}=a_1^{(1)}(k \leftrightarrow p),\nonumber\\
b_1^{(i)}&=&a_1^{(i)}(k \leftrightarrow p),\hspace{20mm}
i=3,4,5,6\nonumber\\
b_1^{(7)}&=&a_1^{(8)}(k \leftrightarrow p),\hspace{30mm}
b_1^{(8)}=a_1^{(7)}(k \leftrightarrow p),\nonumber\\
b_1^{(i)}&=&0,\hspace{38mm}
i=9,10,11,12\nonumber\\
\nonumber\\
b_2^{(1)}&=&a_2^{(2)}(k \leftrightarrow p),\hspace{30mm}
b_2^{(2)}=a_2^{(1)}(k \leftrightarrow p),\;\;\;\;\nonumber\\
b_2^{(i)}&=&a_2^{(i)}(k \leftrightarrow p),\hspace{20mm}
i=5,6\nonumber\\
b_2^{(9)}&=&a_2^{(10)}(k \leftrightarrow p),\hspace{28mm}
b_2^{(10)}=a_2^{(9)}(k \leftrightarrow p),\;\;\;\;\nonumber\\
b_2^{(11)}&=&-a_2^{(12)}(k \leftrightarrow p),\hspace{25mm}
b_2^{(12)}=-a_2^{(11)}(k \leftrightarrow p),\;\;\;\;\nonumber\\
b^{(3)}_{2}&=&(m^2+p^2)(k\cdot p)-(k\cdot p)^2-\frac{p^2}{2}(m^2+k^2),
\nonumber\\
b^{(4)}_{2}&=&m^2(k\cdot p)-\frac{p^2}{2}(m^2+k^2),\nonumber\\
b^{(7)}_{2}&=&m\left[-\Delta^2+\frac{p^2}{2}(p^2-k^2)\right],\nonumber\\
b^{(8)}_{2}&=&m\left[k^2(k\cdot p)-(k\cdot p)^2-2\Delta^2\right]\qquad.
\end{eqnarray}
\newpage
\begin{eqnarray}
c_1^{(i)}&=&0,\hspace{20mm}i=1,..,12\nonumber\\
c^{(1)}_{2}&=&-\frac{p^4k^2}{2}+\frac{3m^4p^2}{2}-m^2p^2k^2+p^2k^2\,k\cdot p
-m^4\,k\cdot p,\nonumber\\
c^{(2)}_{2}&=&-\frac{k^4p^2}{2}-m^2k^2\,k\cdot p+\frac{m^4k^2}{2}
+k^4\,k\cdot p,\nonumber\\
c^{(3)}_{2}&=&(m^2-k^2)\left(\Delta^2+\frac{k^2}{2}(k^2+m^2)
-m^2\,k\cdot p\right),
\nonumber\\
c^{(4)}_{2}&=&\frac{m^4k^2}{2}+k^2p^2\,k\cdot p-m^4\,k\cdot p-\frac{3k^4p^2}{2}
+m^2p^2k^2,\nonumber\\
c^{(5)}_{2}&=&\frac{(k^2-m^2)}{2}\Delta^2,\nonumber\\
c^{(6)}_{2}&=&\frac{m^2(m^2-k^2)}{2}\Delta^2,\nonumber\\
c^{(7)}_{2}&=&m\Bigg[
\left(3(k\cdot p)^2-2p^2\,k\cdot p-k^2\,k\cdot p\right)m^2+
\frac{p^4k^2}{2}+3p^2k^2\,k\cdot p\nonumber\\
&-&2p^2(k\cdot p)^2-\frac{p^2k^4}{2}
-k^2(k\cdot p)^2\Bigg],\nonumber\\
c^{(8)}_{2}&=&m\left[
\left((k\cdot p)^2-2k^2\,k\cdot p+\frac{k^2p^2}{2}+\frac{k^4}{2}\right)m^2
-p^2k^4+p^2k^2\,k\cdot p-k^2(k\cdot p)^2+k^4\,k\cdot p\right],\nonumber\\
c^{(9)}_{2}&=&m\left[
\left(k^2-k\cdot p\right)m^2-\frac{q^2}{2}k^2\right],\nonumber\\
c^{(10)}_{2}&=&m\left[
\left(\frac{3p^2}{2}-2\,k\cdot p+\frac{k^2}{2}\right)m^2
+q^2\,k\cdot p+2\Delta^2\right],\nonumber\\
c^{(11)}_{2}&=&m\left(
m^2q^2\frac{k\cdot p}{2}
+\frac{k^2p^2}{2}(k^2-p^2)+p^2\,k\cdot p\,(k\cdot p\,-k^2)
\right),\nonumber\\
c^{(12)}_{2}&=&m\left[
-m^2\frac{q^2}{2}k^2
-\frac{k^2\,k\cdot p}{2}(p^2+k^2)\right]\qquad.
\end{eqnarray}
\newpage
\begin{eqnarray}
d_1^{(i)}&=&0,\hspace{40mm}i=1,..,12\nonumber\\
d_2^{(1)}&=&c_2^{(2)}(k\leftrightarrow p),\hspace{32mm}
d_2^{(2)}=c_2^{(1)}(k\leftrightarrow p),\;\;\;\nonumber\\
d_2^{(i)}&=&c_2^{(i)},\hspace{37mm}i=5,6\nonumber\\
d_2^{(9)}&=&c_2^{(10)}(k\leftrightarrow p),\hspace{30mm}
d_2^{(10)}=c_2^{(9)}(k\leftrightarrow p),\;\;\;\nonumber\\
d_2^{(11)}&=&-c_2^{(12)}(k\leftrightarrow p),\hspace{27mm}
d_2^{(12)}=-c_2^{(11)}(k\leftrightarrow p),\;\;\;\nonumber\\
d^{(3)}_{2}&=&m^2(k\cdot p)^2-\frac{p^4k^2}{2}+\frac{m^4p^2}{2}
-p^2(k\cdot p)^2
+p^2k^2\,k\cdot p-m^4\,k\cdot p\nonumber\\
d^{(4)}_{2}&=&\frac{m^4p^2}{2}+m^2p^2\,k\cdot p-m^4\,k\cdot p-\frac{k^2p^4}{2},
\nonumber\\
d^{(7)}_{2}&=&m\left[\left(\Delta^2+\frac{p^2}{2}(k^2-p^2)\right)
m^2-p^2\Delta^2\right],\nonumber\\
d^{(8)}_{2}&=&m\left[\left(-k^2\,k\cdot p+(k\cdot p)^2\right)m^2
-p^2\Delta^2-\frac{q^2}{2}k^2p^2\right]\qquad.
\end{eqnarray}
\newpage
\begin{eqnarray}
e^{(1)}_{1}&=&(m^4-p^4)\Delta^2,\nonumber\\
e^{(1)}_{2}&=&(m^4-p^4)\Delta^2+m^2p^4(k^2-p^2),\nonumber\\
e^{(2)}_{1}&=&-2(p^2-m^2)^2(k^2-p^2)\,k\cdot p-(p^4-m^4)\Delta^2,\nonumber\\
e^{(2)}_{2}&=&(m^4-p^4)\Delta^2+m^2p^2k^2(k^2-p^2)
-2m^4\,k\cdot p\,(k^2-p^2),\nonumber\\
e^{(3)}_{1}&=&-(p^4-m^4)\Delta^2+p^2(k^2-p^2)(p^2-m^2)^{2},\nonumber\\
e^{(3)}_{2}&=&(m^4-p^4)\Delta^2-2m^2p^2\,k\cdot p\,(k^2-p^2)
+m^4p^2(k^2-p^2),\nonumber\\
e^{(4)}_{1}&=&-(p^4-m^4)\Delta^2+p^2(k^2-p^2)(p^2-m^2)^2,\nonumber\\
e^{(4)}_{2}&=&(m^4-p^4){\Delta^2}+m^4p^2(k^2-p^2),\nonumber\\
%
%
e^{(5)}_{2}&=&0,\nonumber\\
e^{(6)}_{1}&=&(p^4-m^4)(k^2-p^2)\Delta^2,\nonumber\\
e^{(6)}_{2}&=&(p^4-m^4)(k^2-p^2)\Delta^2,\nonumber\\
e^{(7)}_{1}&=&8mp^2(p^2-m^2)\Delta^2,\nonumber\\
e^{(7)}_{2}&=&m\left[2p^2(p^2-m^2)\Delta^2-m^2p^4(k^2-p^2)\right],\nonumber\\
e^{(8)}_{1}&=&8mp^2(p^2-m^2)\Delta^2,\nonumber\\
%
e^{(8)}_{2}&=&m\left[2p^2(p^2-m^2)\Delta^2+m^2p^2k^2(k^2-p^2)\right],\nonumber\\
%
%
e^{(9)}_{2}&=&m^3(k^2-p^2)\left[p^2-2\,k\cdot p\right],\nonumber\\
%
%
%
e^{(10)}_{2}&=&m^3p^2(k^2-p^2),\nonumber\\
%
%
e^{(11)}_{2}&=&m^3p^2(k^2-p^2)\left[p^2-k\cdot p\right],\nonumber\\
%
%
%
%
%
e^{(12)}_{2}&=&m^3(k^2-p^2)\left[2(k\cdot p)^2-p^2(k\cdot p)-p^2k^2\right],
\nonumber\\
e^{i}_1&=&0,\hspace{20mm}i=5,9,10,11,12\qquad.
\end{eqnarray}
\newpage
\begin{eqnarray}
f_1^{(1)}&=&-e_1^{(2)}(k\leftrightarrow p),\hspace{30mm}
f_1^{(2)}=-e_1^{(1)}(k\leftrightarrow p),\;\;\;\nonumber\\
f_1^{(3)}&=&-e_1^{(4)}(k\leftrightarrow p),\hspace{30mm}
f_1^{(4)}=-e_1^{(3)}(k\leftrightarrow p),\;\;\;\nonumber\\
f_1^{(i)}&=&0,\hspace{42mm}i=5,9,10,11,12\nonumber\\
f_1^{(i)}&=&-e_1^{(i)}(k\leftrightarrow p),\hspace{21mm}i=6,7,8\nonumber\\
f_2^{(1)}&=&-e_2^{(2)}(k\leftrightarrow p),\hspace{30mm}
f_2^{(2)}=-e_2^{(1)}(k\leftrightarrow p),\;\;\;\nonumber\\
f_2^{(3)}&=&-e_2^{(4)}(k\leftrightarrow p),\hspace{30mm}
f_2^{(4)}=-e_2^{(3)}(k\leftrightarrow p),\;\;\nonumber\\
f_2^{(9)}&=&-e_2^{(10)}(k\leftrightarrow p),\hspace{28mm}
f_2^{(10)}=-e_2^{(9)}(k\leftrightarrow p),\;\;\;\nonumber\\
f_2^{(11)}&=&e_2^{(12)}(k\leftrightarrow p),\hspace{31mm}
f_2^{(12)}=e_2^{(11)}(k\leftrightarrow p),\;\;\;\nonumber\\
f_2^{(i)}&=&-e_2^{(i)}(k\leftrightarrow p),\hspace{20mm}i=5,6\nonumber\\
f^{(7)}_{2}&=&m\left[(k^2-p^2)\left(-k^2p^2-2k^2\,k\cdot p
+4(k\cdot p)^2\right)m^2-2k^2(k^2-m^2)\Delta^2\right],\nonumber\\
f^{(8)}_{2}&=&m\left[\left(-2k^2m^2(k^2-p^2)\,k\cdot p-k^4m^2(k^2-p^2)\right)
-2k^2(k^2-m^2)\Delta^2\right]\qquad.\\\nonumber\\
g_1^{(i)}&=&0,\hspace{30mm}i=5,7,8,9,10,11,12\nonumber\\
g_1^{(i)}&=&2\,k\cdot p,\hspace{22mm}i=1,2\nonumber\\
g_1^{(i)}&=&-(k^2+p^2),\hspace{13mm}i=3,4\nonumber\\
g_1^{(6)}&=&-2\,\Delta^2,\nonumber\\
g_2^{(i)}&=&0,\hspace{30mm}i=1,...,12\qquad.\\\nonumber\\
h_1^{(i)}&=&0,\hspace{30mm}i=1,2,4,9,10,11,12\nonumber\\
h_1^{(3)}&=&-4\Delta^2,\nonumber\\
h_1^{(5)}&=&2\Delta^2,\nonumber\\
h_1^{(6)}&=&-2m^2\Delta^2,\nonumber\\
h_1^{(i)}&=&4m\Delta^2,\hspace{21mm}i=7,8\nonumber\\
h_2^{(i)}&=&0,\hspace{30mm}i=5,6,11,12\nonumber\\
h_2^{(1)}&=&h_2^{(2)}(k\leftrightarrow p)=p^2\left[m^2-\,k\cdot p\right]
\Delta^2,\nonumber\\
h_2^{(3)}&=&h_2^{(4)}(k\leftrightarrow p)=k^2p^2-m^2\,k\cdot p,\nonumber\\
h_2^{(7)}&=&h_2^{(8)}(k\leftrightarrow p)=m\left[-p^2\,k\cdot p+\Delta^2
+(k\cdot p)^2\right],\nonumber\\
h_2^{(9)}&=&h_2^{(10)}(k\leftrightarrow p)=m\left[k^2-k\cdot p\right]\qquad.
\end{eqnarray}
\newpage
\newpage
\begin{eqnarray}
l_1^{(1)}&=&l_1^{(2)}(k\leftrightarrow p)=\frac{m^2\Delta^2}{2k^2p^2}
-\left(m^2\frac{(k\cdot p)}{k^2}+p^2\right),\nonumber\\
l_1^{(3)}&=&l_1^{(4)}(k\leftrightarrow p)=
(m^2+k\cdot p)+\frac{m^2\,\Delta^2}{2k^2p^2},\nonumber\\
l_1^{(i)}&=&0,\;\;\;\;\;\;\;\;\;i=5,7,8,9,10,11,12\nonumber\\
l_1^{(6)}&=&\left[-1-\frac{m^2}{2}\left(\frac{1}{p^2}+\frac{1}{k^2}\right)
\right]\Delta^2,\nonumber\\
l_2^{(1)}&=&l_2^{(2)}(k\leftrightarrow p)=
\frac{m^2\Delta^2}{2k^2p^2}-m^2\frac{k\cdot p}{k^2}+p^2,\nonumber\\
l_2^{(3)}&=&l_2^{(4)}(k\leftrightarrow p)=
\frac{m^2\Delta^2}{2\,k^2p^2}+m^2-k\cdot p,\nonumber\\
l_2^{(5)}&=&0,\nonumber\\
l_2^{(6)}&=&\left[1-\frac{m^2}{2}\left(\frac{1}{p^2}+\frac{1}{k^2}\right)
\right]\Delta^2,\nonumber\\
l_2^{(7)}&=&m\left[-k\cdot p+2\frac{(k\cdot p)^2}{k^2}-p^2\right],
\nonumber\\
l_2^{(8)}&=&m\left[-k\cdot p+k^2\right],\nonumber\\
l_2^{(9)}&=&l_2^{(10)}(k\leftrightarrow p)=
m\left[-\frac{k\cdot p}{p^2}+1\right],\nonumber\\
l_2^{(11)}&=&-l_2^{(12)}(k\leftrightarrow p)=
m\left[\frac{(k\cdot p)^2}{k^2}-p^2\right]\qquad.
\end{eqnarray}

\end{document}